\documentclass[a4paper,11pt,twoside]{book}
\usepackage{pdfpages}

\usepackage{lipsum} 
\usepackage[english]{babel}
\usepackage[scale=0.75]{geometry} 
\usepackage{nccmath}
\usepackage{amsmath}
\usepackage{nccmath}
\usepackage{subcaption}
\usepackage{amsfonts}
\usepackage{import}

\newcommand{\Df}{{\Delta_\phi}}
\newcommand{\tr}{\mathrm{tr}}

\usepackage{setspace}
\usepackage[T1]{fontenc}
\usepackage[utf8]{inputenc}
\usepackage{lmodern}
\usepackage[english]{babel}
\usepackage{textgreek}

\usepackage{fixltx2e}

\usepackage{geometry} 
\reversemarginpar

\usepackage{amsmath,amssymb,amsfonts,amsthm}
\usepackage{mathrsfs}

\DeclareMathOperator{\e}{e}

\usepackage[locale = FR]{siunitx} 
\DeclareSIUnit\vitesse{\meter\per\second}
\usepackage{eurosym}
\DeclareSIUnit{\octet}{o}

\usepackage{graphicx,array,tikz,multirow}
\usepackage{caption,subcaption} 
\usepackage{svg,float}
\usepackage{booktabs,paralist}
\newcolumntype{x}[1]{>{\centering\arraybackslash\hspace{0pt}}p{#1}}
\usepackage[section]{placeins}
\usepackage{hanging}

\usepackage{fancyhdr,emptypage} 
\let\cleardoublepage\clearpage
 
\fancypagestyle{plain}{ 
    \fancyhead{}\fancyfoot[C]{\thepage}

}

\usepackage[Lenny]{fncychap}
\ChNameVar{\fontsize{25}{25}\usefont{OT1}{phv}{m}{n}\selectfont}
\ChRuleWidth{0pt}
\ChNumVar{\fontsize{60}{62}\selectfont\textcolor{curcolor}}

\makeatletter
\ChTitleVar{\Huge\rm}
\renewcommand{\DOCH}{%
\setlength{\fboxrule}{\RW} 
\fbox{\CNV\FmN{\@chapapp}\space \CNoV\thechapter}\par\nobreak
\vskip 20\p@}
\renewcommand{\DOTIS}[1]{%
\CTV\bfseries\FmTi{#1}\par\nobreak
\vskip 20\p@}
\makeatother

\usepackage[francais,nohints,tight]{minitoc}		
\dominitoc

\usepackage[nottoc]{tocbibind} 
\usepackage[titles]{tocloft}

\usepackage{textcomp} 
\usepackage{xcolor} 
\usepackage{epigraph} 
\usepackage{titling}
\usepackage{lipsum} 
\usepackage{csquotes} 

\usepackage{xspace}
\usepackage{afterpage}

\usepackage[textsize=footnotesize]{todonotes}

\usepackage{bookmark}
\usepackage{acronym}
\usepackage[nameinlink,french]{cleveref} 
\Crefname{figure}{Fig.}{Figs.} 
\crefname{figure}{fig.}{figs.}
\Crefname{equation}{Eq.}{Eqs.}
\crefname{equation}{eq.}{eqs.}
\Crefname{table}{Table.}{Tables.}
\crefname{table}{table.}{tables.}

\definecolor{color_ref}{rgb}{0.18, 0.31, 0.31} 
\definecolor{color_link}{RGB}{36, 56, 141}
\definecolor{curcolor}{RGB}{113,127,184} 

\hypersetup{
	colorlinks=true, 
	pdfstartview=FitV, 
	urlcolor=color_link, 
	linkcolor= color_link, 
	citecolor=color_ref 
}

\usepackage[hyperref=true,natbib=true,
			backref=true,date=year,
			backend=biber,
			url=false,doi=false,isbn=false,%
			minbibnames=6, 
			maxbibnames=6, 
			maxcitenames=1,mincitenames=1, 
			maxalphanames=1, 
			style=alphabetic,
			sorting=nyt]{biblatex}

\renewbibmacro{in:}{} 

\DeclareLabelalphaTemplate{
  \labelelement{
    \field[final]{shorthand}
    \field{labelname}
    \field{label}
  }
  \labelelement{\literal{,\addhighpenspace}}
  \labelelement{\field{year}}
}

\AtEveryBibitem{%
    \clearfield{note} 
    \clearlist{language} 
}
\newcommand{\sujet}[1]{\renewcommand{\sujet}{#1}}
\newcommand{\auteur}[1]{\renewcommand{\auteur}{#1}}
\newcommand{\encadrant}[1]{\renewcommand{\encadrant}{#1}}

\newcommand{\addchapnonumber}[1]{
\phantomsection
\addtocounter{chapter}{1}
\chapter*{#1}
\addcontentsline{toc}{chapter}{#1}
\markboth{#1}{#1}
\setcounter{section}{0}
}

\newcommand{\AddResumeAbstract}{
\chapter{Résumé}

\thefrabstract
\vskip 2em \noindent\makebox[\linewidth]{\rule{.5\linewidth}{0.4pt}}
\vfill
\noindent 

\chapter{Abstract}

\theenabstract
\vskip 2em \noindent\makebox[\linewidth]{\rule{.5\linewidth}{0.4pt}}
\vfill
\noindent
} 
\usepackage[acronym,toc,nogroupskip,section=chapter,nonumberlist,nopostdot]{glossaries}
\makeglossaries

\newcommand{\newacronymen}[4]{\newacronym{#1}{#2}{\textit{#3} (#4)}}

\newacronymen{gpu}{GPU}{Graphical Processing Unit}{processeur graphique}
\newacronym{ram}{RAM}{\textit{Random Access Memory}}
\newacronym{ssd}{SSD}{\textit{Solid-State Drive}}
\usepackage[nomaketitle]{config/psl-cover/psl-cover}

\sujet{Méthode bootstrap en physique théorique
\\Bootstrap Method in Theoretical Physics}
\author{Zechuan Zheng}
\encadrant{Directeur \textsc{De Recherche}}

\institute{École normale supérieure - PSL}
\doctoralschool{Physique en Île-de-France}{564}
\specialty{Physique des hautes éner- gies - Théorie}
\date{15/09/2023}

\pslassetspath{config/psl-cover}

\jurymember{2}{Alice \textsc{Guionnet}}{Directeur de recherche, École normale supérieure de Lyon}{Président} 
\jurymember{3}{Alexander \textsc{Zhiboedov}}{Chercheur permanent senior, CERN-Organisation européenne pour la recherche nucléaire}{Rapporteur}
\jurymember{4}{Agnese \textsc{Bissi}}{Associate professor, Uppsala University}{Rapporteur}
\jurymember{5}{Alexander \textsc{Migdal}}{Professeur, New York University}{Examinateur}
\jurymember{6}{Miguel \textsc{Paulos}}{Chargé de recherche, École Normale Supérieure}{CoDirecteur de thèse}
\jurymember{7}{Vladimir \textsc{Kazakov}}{Professeur, École Normale Supérieure}{Directeur de thèse}

\frabstract{
Dans le domaine de la physique contemporaine, la méthode bootstrap est généralement associée à une approche basée sur l’optimisation pour résoudre des problèmes. Cette méthode exploite notre compréhension d’un problème physique spécifique, utilisée comme les contraintes pour le problème d’optimisation, afin de délimiter la région autorisée de notre théorie physique. Notamment, cette méthode donne souvent non seulement des limites numériques précises pour les quantités physiques mais offre également des perspectives théoriques sur la nature du problème en question. La méthode numérique bootstrap moderne a connu son plus grand succès dans les domaines de la théorie des champs conformes (via le bootstrap conforme) et de l’amplitude de dispersion (à travers le bootstrap de la matrice S). Cette thèse présente l’application de la méthode bootstrap aux modèles matriciels (matrices aléatoires), à la théorie de Yang-Mills et à la théorie des champs conformes. Nous commencerons par un examen des éléments fondamentaux de ces théories. Ensuite, nous nous plongerons dans les études bootstrap de ces modèles.

Nous proposons une méthode bootstrap de relaxation améliorée pour la résolution numérique de modèles multi-matriciels à la limite de N grand. Cette méthode fournit des inégalités rigoureuses sur les moments de trace simple des matrices jusqu’à un ordre “coupure” spécifié des moments. Nous avons une preuve rigoureuse de l’applicabilité de cette méthode dans le cas du modèle à une matrice. Nous démontrons l’efficacité numérique de notre méthode en résolvant le modèle à deux matrices analytiquement “insolvable” avec une interaction tr[A, B]² et des potentiels quartiques.
Nous étendons ensuite notre étude à la théorie de Yang-Mills sur réseau aux dimensions 2, 3 et 4 en utilisant la méthode numérique bootstrap. Notre approche combine des équations de boucle, avec une coupure sur la longueur maximale des boucles, et des conditions de positivité sur certaines matrices de moyennes de boucles de Wilson. Nos résultats suggèrent que cette approche bootstrap peut offrir une alternative tangible à l’approche de Monte Carlo, jusqu’à présent incontestée.

Nous explorons le problème de la mise en limite des corrélateurs CFT dans la section euclidienne. En reformulant la question comme un problème d’optimisation, nous construisons numériquement des fonctionnelles qui déterminent des limites supérieures et inférieures sur les corrélateurs dans plusieurs circonstances. Notre analyse révèle que le corrélateur d’Ising 3d prend les valeurs minimales possibles autorisées dans la section euclidienne. Nous découvrons également un intrigant CFT 3d qui sature les limites de l’écart, de la maximisation OPE et des valeurs des corrélateurs.}

\enabstract{
In the realm of contemporary physics, the bootstrap method is typically associated with an optimization-based approach to problem-solving. This method leverages our understanding of a specific physical problem, which is used as the constraints for the optimization problem, to carve out the allowed region of our physical theory. Notably, this method often yields not only precise numerical bounds for physical quantities but also offers theoretical insights into the nature of the problem at hand. The modern numerical bootstrap method has seen its greatest success in the fields of conformal field theory (via the conformal bootstrap) and Scattering amplitude (through the S-matrix bootstrap). This dissertation presents the application of the bootstrap method to matrix models (random matrices), Yang-Mills theory, and conformal field theory. We will commence with a review of the fundamental elements of these theories. Following this, we will delve into the bootstrap studies of these models. 

We propose an enhanced relaxation bootstrap method for the numerical resolution of multi-matrix models in the large N limit. This method provides rigorous inequalities on the single trace moments of the matrices up to a specified “cutoff” order of the moments.  We have rigorous proof of the applicability of this method in the case of the one-matrix model. We demonstrate the numerical efficiency of our method by solving the analytically “unsolvable” two-matrix model with tr[A, B]² interaction and quartic potentials. We further extend our study to the lattice Yang-Mills theory at dimensions 2, 3, and 4 using the numerical bootstrap method. Our approach combines loop equations, with a cutoff on the maximal length of loops, and positivity conditions on certain matrices of Wilson loop averages. Our results suggest that this bootstrap approach can provide a tangible alternative to the, so far uncontested, Monte Carlo approach.

We explore the problem of bounding CFT correlators in the Euclidean section. By reformulating the question as an optimization problem, we construct functionals numerically which determine upper and lower bounds on correlators under several circumstances. Our analysis reveals that the 3d Ising correlator takes the minimal possible allowed values on the Euclidean section. We also uncover an intriguing 3d CFT that saturates gap, OPE maximization, and correlator value bounds.
}

\frkeywords{Méthode bootstrap, Modeles des matrices, Théorie des champs conforme}
\enkeywords{Bootstrap Method, Matrix models, Conformal field theory}

\title{\sujet}
\hypersetup{
pdftitle={Thesis Year},
pdfsubject={\sujet},
pdfauthor={\theauthor},
pdfkeywords={\thefrkeywords},
}
\begin{document}
\pslcover{} 

\frontmatter
\chapter{Acknowledgement}

I would like to express my deepest gratitude and appreciation to my supervisor, Vladimir Kazakov, for his invaluable guidance, unwavering support, and encouragement throughout the course of my PhD journey. His knowledge and experience in the field, combined with his patience, have been instrumental in the completion of this work. I am privileged to have been one of his students, and his mentorship has not only shaped this thesis but has also significantly impacted my personal and professional development. I have learned so much from his wisdom, and his intellectual rigor has pushed me to broaden my thinking and approach to research.

In the same vein, I would also like to express my sincere thanks to my co-supervisor, Miguel Paulos. His insightful comments, constructive criticisms, and detailed feedback have greatly enhanced the quality of this thesis. His unwavering faith in my abilities even during tough times has been a constant source of motivation. His enthusiasm for research and dedication to teaching have left an indelible impression on me.

In addition, I would like to express my profound gratitude to Edouard Brézin, Davide Gaiotto, Slava Rychkov,  Pedro Vieira and Xi Yin. I am deeply appreciative of the essential role they have played in the progression of my academic journey, particularly their assistance and guidance during various application processes. Their faith in my potential and their willingness to support my professional endeavors have been a significant contribution to the progression of my academic journey.

I am sincerely thankful to Agnese Bissi, Alice Guionnet, Alexander Migdal, and Alexander Zhiboedov for accepting the responsibility of being part of the jury for my thesis defense. Their presence and valuable insights added significant depth to my work and their constructive feedback has contributed to the quality of this thesis.

I would also like to express my sincere gratitude to my esteemed colleagues in the global physics research community. Your guidance, collaborations, and shared intellectual curiosity have significantly enriched my research journey. Specifically, I would like to acknowledge António Antunes, Camille Aron, Avik Banerjee, Benjamin Basso, Lior Benizri, Antoine Bourget, Xiangyu Cao, Xiaowen Chen, Zhongwu Chen, Sihao Cheng, Minjae Cho, Zhihao Duan, David Jaramillo Duque, Manuel Díaz, Nikolay Ebel,  Rajeev Erramilli, Farah El Fakih, Lucija Farkaš, Barak Gabai, Kausik Ghosh, Gabriel Gouraud, Yinchen He, Yifei He, Aditya Hebbar, Ludwig Hruza, Hongye Hu, Yangrui Hu, Amir-Kian Kashani-Poor, Apratim Kaviraj, Arthur Klemenchuk Sueiro, Ivan Kostov, Werner Krauth, Augustin Lafay, Ho Tat Lam, Bingxin Lao, Nat Levine, Songyuan Li, Wenliang Li, Yuezhou Li, Zhijin Li, Henry Lin, Ying-Hsuan Lin, Ruochen Ma, Marina Krstic Marinkovic, Dalimil Mazac, Maria Neuzil, Colin Oscar Nancarrow, Enrico Olivucci, Vassilis Papadopoulos, Sabrina Pasterski, Joao Penedones, Eric Perlmutter, Giuseppe Policastro, Jiaxin Qiao, Mingfeng Qiu, Marten Reehorst, Balt van Rees, Victor Rodriguez, Junchen Rong, Joshua Sandor, Didina Serban, Benoit Sirois, Evgeny Sobko, Xueyang Song, Ning Su, Tin Sulejmanpasic, Zimo Sun, Petar Tadić, Emilio Trevisani, Jan Troost, Alexander Tumanov, Alessandro Vichi, Xing Wang, Chong Wang, Zixia Wei, Yuan Xin, Hari Yadalam, Kang Yang, Weicheng Ye, Zahra Zahraee, Bernardo Zan, Yi Zhang, Xiang Zhao, Deliang Zhong, and Liujun Zou. Your contributions to the academic community, intellectual discussions, and encouragement have played a crucial role in my growth as a researcher. I am truly grateful for your support and am excited about our continued collaboration in furthering the frontiers of physics.

To my friends at École Normale Supérieure, I owe a deep sense of gratitude. Your support, companionship, and the shared experiences have been pivotal in this journey. Specifically, I would like to acknowledge Yawen Chen, Qiyuan Chen, Jinhui Cheng, Senwen Deng, Wenshuo Fan, Jiahui Feng, Hao Fu, Tiancheng He, Yue Hong, Chuhao Huang, Mengzi Huang, Yulai Huang, Gang Lin, Haohao Liu, Youmin Liu, Daheng Min, Junhui Qin, Liang Qin, Yue Qi, Yi Shan, Qing Wang, Songbo Wang, Yitong Wang, Keyu Wang, Zhenhong Wang, Ziao Wang, Yijun Wan, Shiyi Wei, Baojun Wu, Yuxiao Xie, Minchen Xia, Ziyu You, Yilin Ye, Yiwei Zhang, Chao Zhang, Yiru Zheng, Qimeng Zhu and Zhaoxuan Zhu. Despite my attempt to be comprehensive, I apologize if anyone has been inadvertently left out. Know that your influence and contribution to my life and work at ENS are deeply appreciated.

Lastly, and most importantly, I would like to express my deepest gratitude to my parents, who are thousands of miles away in China. Despite the distance and the challenges posed by the COVID-19 pandemic which has kept us apart, their unwavering support, encouragement, and love have been my source of strength and resilience. The pandemic has reminded us of the fragility of life and the importance of family. Even though I have been unable to return home and see them, their constant care, understanding, and faith in my abilities have been a comforting presence throughout this journey. They have been my rock, providing me with the courage and determination to overcome the obstacles I have faced.
\AddResumeAbstract 

\tableofcontents\newpage
\mainmatter
\setcounter{page}{1}
\chapter{Introduction}\label{chp1_intro}

\section{Background}

In the realm of modern quantum field theory (QFT), the importance of symmetry cannot be overstated. Our progressive understanding of QFT has invariably been coupled with insights gleaned from symmetries, whether assumed (like Poincaré symmetry and the constraining power of anomalies), or spontaneously broken.

\begin{figure}[h!]
\centering
\includegraphics[scale=.25]{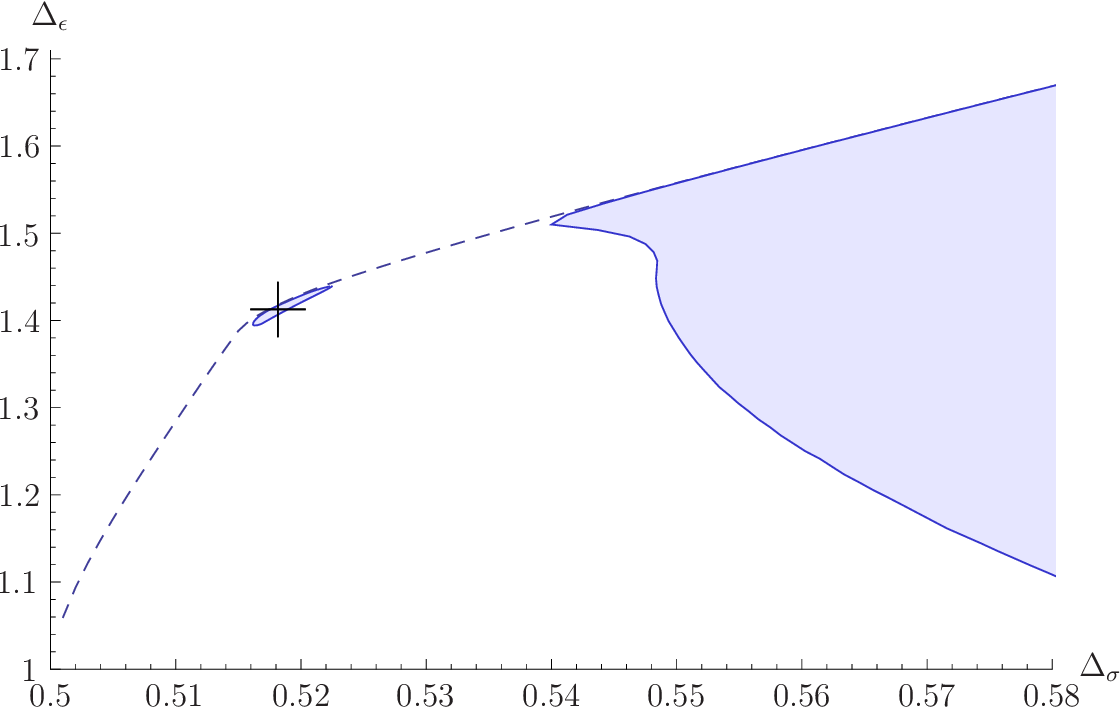}
\caption{The depicted region corresponds to the parameter space allowed by the $3d$ Ising model. It is defined by the constraints from three distinct four-point correlators, under the assumption that there is only a single relevant $\mathbb{Z}_2$-odd operator present. The plot has been adapted from the work by Kos et al. \cite{Kos:2014bka}.}
\label{fig:ising}
\end{figure}

The bootstrap philosophy in physics embodies a methodology that aims to exploit the potential power of symmetry to its utmost limit. The approach seeks to solve problems primarily using the assumption of symmetry, along with other generic constraints such as unitarity and locality. This method eschews reliance on the microscopic details of the problem at hand. Although this perspective was initially proposed in the 1960s within the context of the S-matrix \cite{Chew:110879}, it was largely overshadowed due to the subsequent successes in Quantum Chromodynamics (QCD).

The first significant outcome following the bootstrap philosophy did not materialize until the 1980s, when Belavin, Polyakov, and Zamolodchikov (BPZ) used the "conformal bootstrap" to solve minimal models with $c<1$ \cite{belavin1984infinite}. Their solution relied heavily on the infinite symmetries present in $2d$ conformal field theory, a feature absent in other dimensions.

The 21st century has seen an unprecedented surge in the development of mathematical optimization theory, largely spurred by the rapid expansion of the machine learning industry, which had a significant breakthrough in 2006 \cite{10.1162/neco.2006.18.7.1527}. Around the same period, a seminal work emerged, which used optimization theory to bound the dynamics of $4d$ conformal field theory (CFT) \cite{Rattazzi:2008pe}. This method, known as the ``bootstrap method,'' sparked renewed interest in the application of optimization theory to tackle problems in theoretical physics \cite{Ferrara:1973yt, 1974JETP...39...10P}. Despite its numerical nature, the bootstrap method has proven successful in generating rigorous bounds for dynamical quantities. The approach has seen particularly noteworthy success within the context of CFT \footnote{Comprehensive reviews of these advancements can be found in \cite{Poland:2018epd, Bissi:2022mrs}. See also \cite{Poland:2022qrs, Hartman:2022zik} for more recent updates. }, demonstrating that very accurate numerical results can be obtained solely from assumptions of CFT axioms, unitarity, and global symmetries.

Conformal invariance and the operator product expansion led to the so-called bootstrap equation:
\begin{equation}\label{boot}
\sum_{\Delta} a_{\Delta}^2 F_{\Delta}\left(z,\bar{z}\right)=0,\quad (z,\bar{z})\in (\mathbb{C}\backslash (-\infty,0)\cup(1,\infty))^2
\end{equation}
Here, conformal invariance helps us determine $F_{\Delta}[z,\bar{z}]$ (either analytically or numerically), and unitarity gives us the reality and potential $\Delta$ region in Eq. (\ref{boot}).

The aforementioned situation embodies an infinite generalization of linear programming. From this, we can obtain bounds for intriguing dynamical quantities such as the dimension gap above the identity operator. For instance, consider a functional $\Lambda$ acting on $F_{\Delta}$, where $\Lambda(F_{\Delta})\geq 0$ for $\Delta\geq \Delta_{\mathrm{gap}}$ and $\Lambda(F_{0})> 0$. The existence of such a functional, coupled with Eq. (\ref{boot}), implies that there must be at least one operator in the region $0<\Delta<\Delta_{\mathrm{gap}}$. By searching for $\Lambda$ across all possible $\Delta_{\mathrm{gap}}$, we can obtain an optimal bound for the dimension gap.

Numerous subsequent studies utilizing the bootstrap approach have yielded compelling results, showcasing the effectiveness of this methodology. One of the most significant achievements in this vein has been the determination of the operator dimensions in the 3D Critical Ising model with unprecedented precision \cite{El-Showk:2012cjh, El-Showk:2014dwa, Kos:2014bka}. As depicted in Fig. \ref{fig:ising}, the $3d$ critical Ising model, as determined by the conformal bootstrap, resides within a minuscule island in the parameter space.

\begin{figure}[h]
    \centering
    \includegraphics[width=0.45\textwidth]{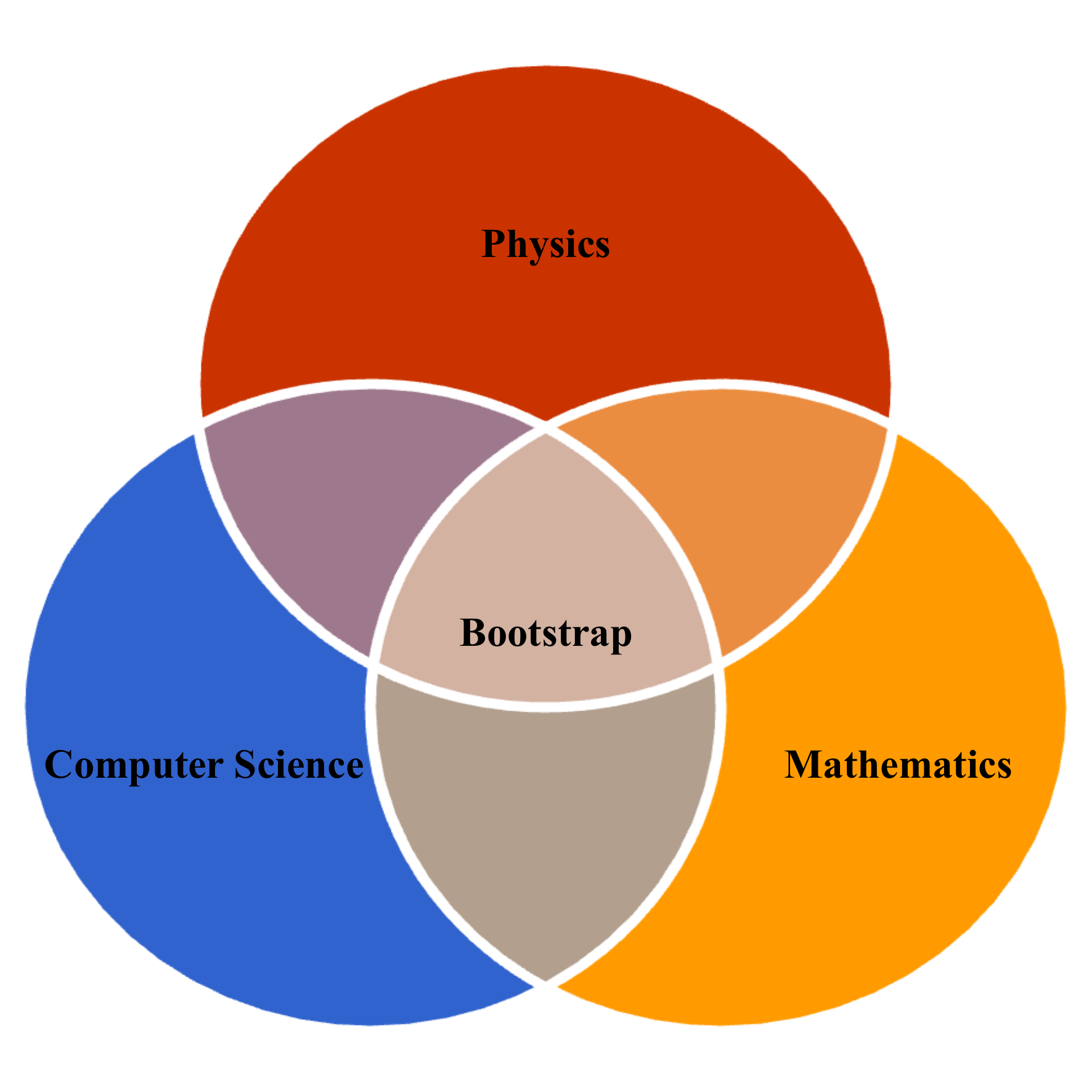}
    \caption{The bootstrap method.}
    \label{fig: relationship}
\end{figure}

The successful execution of the bootstrap method, showcased in Figure~\ref{fig: relationship}, is a testament to the essential interplay between physics, mathematics, and computer science:

\begin{itemize}
\item On the physics front, a comprehensive understanding of the underlying principles is key for devising effective constraints. Notable instances of this include the application of the conformal block expansion\cite{Dolan:2003hv} of the four-point correlator in conformal field theory, and the usage of Makeenko-Migdal loop equations\cite{Makeenko:1979pb} in gauge theory as non-trivial constraints within bootstrap calculations.

\item Mathematics lends the necessary toolset to interpret and validate the constraints associated with the physical model. For instance, solid mathematical analysis is required to ascertain the numerical veracity of the conformal bootstrap \cite{Pappadopulo:2012jk}.

\item In the sphere of computer science, robust techniques are vital for the practical implementation of the bootstrap method. Proficiency in programming is necessitated for the development of specialized optimization solvers, such as \textit{SDPB} \cite{Simmons-Duffin:2015qma, Reehorst:2021ykw, Liu:2023elz}, \textit{JuliBoots} \cite{2014arXiv1412.4127P}, \textit{PyCFTBoot}\cite{Behan:2016dtz} and \textit{FunBoot} \cite{Ghosh:2023onl}, which are designed specifically to tackle bootstrap problems. It is also noteworthy that substantial numerical outcomes from the bootstrap method are typically generated using supercomputing resources \cite{Chester:2019ifh}.
\end{itemize}

Despite the impressive advancements in conformal bootstrap, the application of the bootstrap method to systems devoid of the benefit of conformal symmetry remains a challenging problem. A potential area of exploration is the S-matrix bootstrap \cite{Homrich:2019cbt, Kruczenski:2022lot, Paulos:2016but, Paulos:2016fap, Paulos:2017fhb}, where the author embarked on some preliminary investigation into this area during his master's studies \cite{Paulos:2018fym}. However, the progress in the field of S-matrix bootstrap has largely been confined to the establishment of broad bounds on the parameters of the theory. As of now, it is not feasible to effectively incorporate information or constraints derived from the ultraviolet (UV) theory into the S-matrix bootstrap framework.

In light of recent advancements \cite{Anderson:2016rcw, Lin:2020mme, Han:2020bkb}, a novel bootstrap approach has emerged for investigating matrix models. This method is characterized by not only adhering to general assumptions such as unitarity and global symmetries but also by incorporating relations among physical observables imposed by the equations of motion.\footnote{For a detailed study on the mathematical convergences of this method, readers are directed to \cite{2022arXiv221005239G, Kazakov:2021lel, Cho:2023ulr}.} Its wide applicability has been quickly demonstrated in lattice field theories \cite{Anderson:2016rcw, Anderson:2018xuq, Cho:2022lcj, Kazakov:2022xuh}, matrix models \cite{Han:2020bkb, Jevicki:1982jj, Jevicki:1983wu, Koch:2021yeb, Lin:2020mme, Lin:2023owt, Mathaba:2023non}, quantum systems\cite{Aikawa:2021eai, Aikawa:2021qbl, Bai:2022yfv, Berenstein:2021dyf, Berenstein:2021loy, Berenstein:2022unr, Berenstein:2022ygg, Berenstein:2023ppj, Bhattacharya:2021btd, Blacker:2022szo, Ding:2023gxu, Du:2021hfw, Eisert:2023hcx, Fawzi:2023ajw, hanQuantumManybodyBootstrap2020, Hastings:2021ygw, Hastings:2022xzx, Hessam:2021byc, Hu:2022keu, Khan:2022uyz, Kull:2022wof, Li:2022prn, Li:2023nip, Morita:2022zuy, Nakayama:2022ahr, Nancarrow:2022wdr, Tavakoli:2023cdt, Tchoumakov:2021mnh, Fan:2023bld, Fan:2023tlh, John:2023him, Li:2023ewe, Zeng:2023jek}, and even classical dynamical systems \cite{goluskinBoundingAveragesRigorously2018, goluskinBoundingExtremaGlobal2020, tobascoOptimalBoundsExtremal2018, Cho:2023xxx}.


The initial part of this thesis discusses our application of this novel bootstrap method to matrix models \cite{Kazakov:2021lel} and large $N$ lattice gauge theories \cite{Kazakov:2022xuh}. In our first publication, we rigorously justified the bootstrap approach within the context of a one-matrix model and introduce a relaxation on the quadratic terms in the loop equations, thereby attaining unprecedented precision in an unsolvable two-matrix model. In our second publication, we further expand the application of this bootstrap method (along with the relevant relaxation) to the large $N$ lattice gauge theory. Remarkably, this approach yielded excellent results in the estimation (or more precisely, the bounding) of the plaquette average of the lattice gauge theory.

The latter portion of this thesis delves into the author's collaborative work with Miguel Paulos, focusing on bounding correlation functions via the conformal bootstrap\cite{Paulos:2021jxx}. Given the unique exponential behavior of the target function in this specific bootstrap problem, the traditional polynomial approximation used in \textit{SDPB}\cite{Simmons-Duffin:2015qma} is unsuitable. We demonstrated that the solution for maximizing the gap usually saturates the minimization of the correlator. Meanwhile, the maximization of the correlator, subject to additional constraints, effectively reproduces the Mean-field theory.

In the final chapter of this thesis, we consolidate our findings, offering a concise summary and implications of our work. Moreover, we suggest a variety of promising directions for future research and exploration.


\section{Main results}

\subsection{Matrix bootstrap}

A major portion of my research agenda revolves around the matrix bootstrap \cite{Kazakov:2021lel}. This endeavor includes two crucial components: a solid justification of the bootstrap method for the one-matrix model using the results of the Hamburger moment problem, and a noteworthy advancement of this method for bootstrapping large $N$ matrix models via convex relaxation.

More specifically, the justification process, capitalizing on the results of the Hamburger moment problem, rigorously affirms that the positivity of the correlation matrix implies the positivity of the resolvent, and vice versa:

\begin{equation}\label{equiv}
    \textit{Positivity of correlation matrix}\Leftrightarrow\textit{Positivity of Resolvent}
\end{equation}

This equivalence was employed in conjunction with the loop equations to analytically and exhaustively categorize the solutions of the large $N$ matrix model specified below:

\begin{equation}\label{MMint}
    Z_{N}=\int d^{N^2}M\,\mathrm{e}^{-N\mathrm{tr} { V}(M)} , \quad V(x)=\frac{1}{2}\mu x^2 +\frac{1}{4}g x^4
\end{equation}

The latter half of the paper focuses on the resolution of the subsequent model:

\begin{equation}\label{2MMcom21}
Z=\lim_{N\rightarrow \infty}\int d^{N^2}A,d^{N^2}B,\mathrm{e}^{-N \mathrm{tr}\left( -h[A,B]^2/2+A^2/2+g A^4/4+B^2/2+g B^4/4\right)}
\end{equation}

Given our existing methodologies, this model lacks an analytical solution. To bootstrap this model, we employed a relaxation method as follows:

\begin{equation}\label{relaxm}
    Q=x x^\mathrm{T}\Rightarrow\mathcal{R}=\begin{pmatrix}
1 & x^{\mathrm{T}}\\
x & Q
\end{pmatrix}\succeq 0.
\end{equation}

\begin{figure}
\centering
\includegraphics[scale=.65]{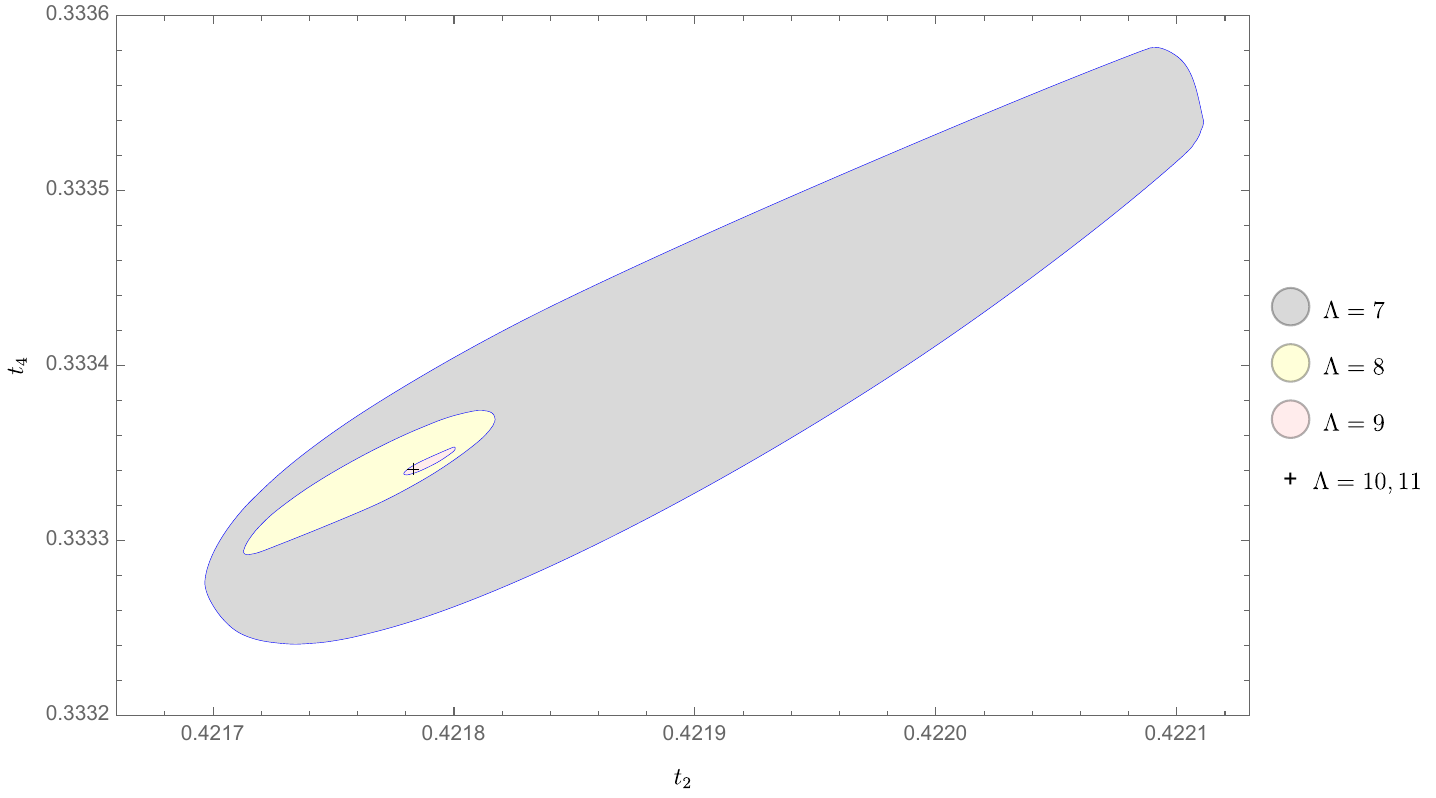}
\caption{The allowed region of \(t_2-t_4\) of model~\eqref{2MMcom21} with parameter $g=1, h=1$ for the cutoff  \(\Lambda=7,8,9,10,11.\) We recall the definition of \(\Lambda\): the longest operators in the correlation matrix and in the loop equations have the length $2\Lambda$.}
\label{fig: covg}
\end{figure}

In the above, $Q$ represents the quadratic terms in the loop equations, derived from the large $N$ factorization. The vector $x$ refers to the column vector of single trace variables. The results obtained from the relaxed bootstrap problem considerably outshine the precision and efficiency of the conventional numerical method for large $N$ matrix models, the Monte Carlo (MC) method \cite{Jha:2021exo}. For $g=h=1$, we achieved a 6-digit precision result:

\begin{equation}
    \begin{cases}
    0.421783612\leq t_2 \leq 0.421784687\\
    0.333341358\leq t_4 \leq 0.333342131
    \end{cases}
\end{equation}

The illustration of the contraction of the permissible domain as a function of the corresponding bootstrap cutoff is visually demonstrated in Figure~\ref{fig: covg}.

\subsection{Bootstrapping the Lattice Yang-Mills Theory}
Another project \cite{Kazakov:2022xuh} involved using this method to bootstrap the single plaquette Wilson loop average in large $N$ lattice Yang-Mills theory.\footnote{For other bootstrap studies of the gauge theory or QCD, the reader could refer to \cite{Nakayama:2014sba, Albert:2022oes, Albert:2023jtd, Fernandez:2022kzi, Guerrieri:2018uew, He:2023lyy, Caron-Huot:2023tpw, Ma:2023vgc}. It is a widespread belief that large N Yang-Mills is equivalent to large N QCD because of the suppression of the fermionic component in the planar limit. However, recent evidence suggests that this assumption may be more nuanced~\cite{Cherman:2022eml}.}  The results are encouraging when compared with the MC method, especially considering that the MC method for lattice QCD has undergone intense investigation over several decades. 

\begin{figure}[h]
\centering
    \includegraphics[width=.75\textwidth]{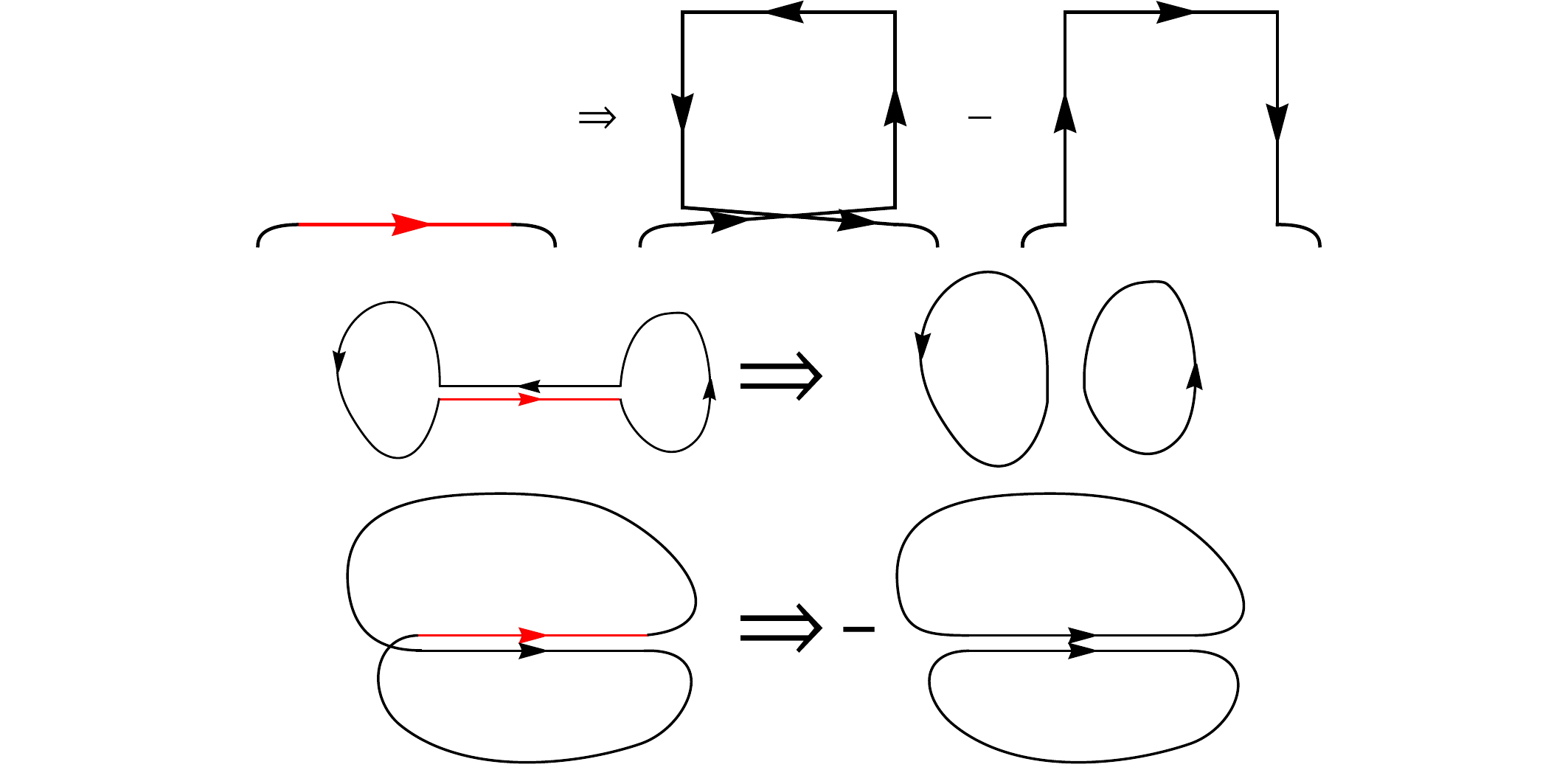}
    \caption{Schematic representation of LEs: The first line shows the variation of a link of Wilson loop in the LHS of Eq.\eqref{MMLE}. The 2nd and 3rd lines show the splitting of the contour along the varied line into two sub-contours, for two different orientations of coinciding links in the RHS of Eq.\eqref{MMLE}.}
 \label{fig:LE}   
\end{figure}

We analyze the Wilson-action-based Large Gauge Theory (LGT)~\cite{Wilson:1974sk} represented by $S=-\frac{N_c}{\lambda}\sum_{P} {\mathrm{Re}}\,\mathrm{tr} U_P$, with $U_P$ denoting the product of four unitary link variables around the plaquette $P$. Considering the 't~Hooft limit ($N_c\to\infty$), the primary focus lies on the Wilson loop Averages (WAs): $W[C]=\langle\frac{\mathrm{tr}}{N_c}\prod_{l\in C} U_l\rangle$, where the matrix product traverses link variables within the lattice loop $C$. WAs adhere to Makeenko-Migdal Loop Equations (LEs)\cite{Makeenko:1979pb}, or Schwinger-Dyson equations, which encapsulate the measure invariance concerning group shifts, such that $U_l\to U_l(1+i\epsilon)$.\footnote{The recent advancement involves utilizing loop equations to address issues in turbulence.\cite{Migdal:2022bka, Migdal:2023ppb}.} The LEs are represented schematically:

\begin{equation}\label{MMLE}
\sum_{\nu\perp\mu}\left(W[C_{l_\mu}\!\!\cdot\overrightarrow{\delta C^{\nu}_{l_\mu}}]-W[C_{l_\mu}\!\!\cdot\overleftarrow{\delta C^{\nu}_{l_\mu}}]\right)=\lambda\sum_{\underset{l^\prime\sim l}{l^\prime\in C}}\,\epsilon _{ll^\prime}W[C_{ll^\prime}]\,\,W[C_{{l^\prime l}}]
\end{equation}
with the LHS indicating the loop operator's action on the link $l_\mu$, and the RHS signifying the contour splitting $C\to C_{ll^\prime}\cdot C_{l^\prime l}$, as shown in Figure~\ref{fig:LE}.

\begin{figure}
\centering
\includegraphics[width=.75\textwidth]{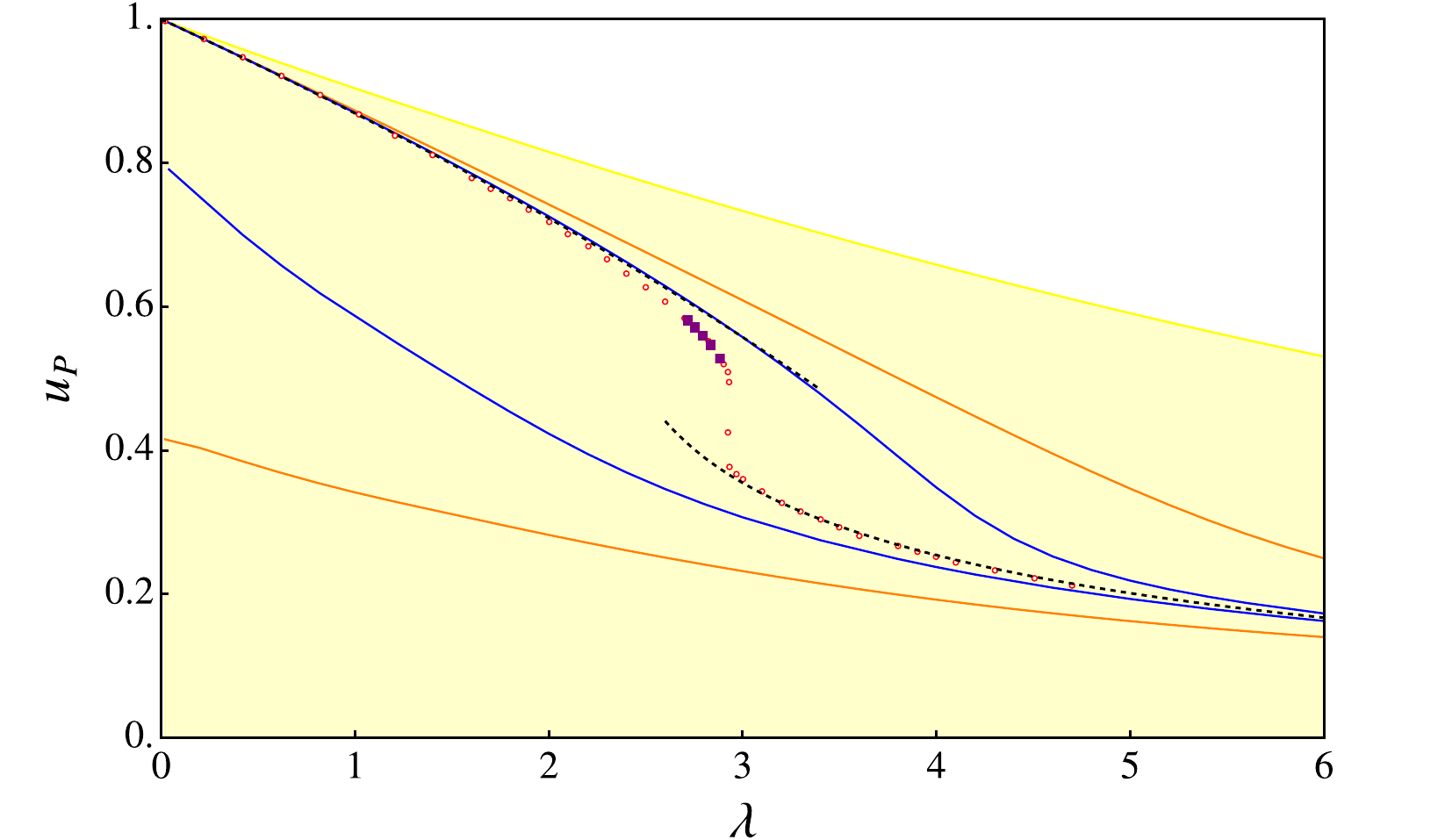}
\caption{The figure presents our bootstrap results for upper and lower bounds on the plaquette average in \(4D\) LGT. The domains for \(L_{\mathrm{max}}=8, 12, 16\) are respectively depicted in yellow, orange, and blue. Red circles represent the Monte Carlo (MC) data for \(SU(10)\) LGT, with 5 purple squares indicating the results for \(SU(12)\). The upper and lower dashed lines signify the 3-loop perturbation theory~\cite{Alles:1998is} and strong coupling expansion~\cite{Drouffe:1983fv}, respectively.}
\label{fig:plaquette4D}
\end{figure}

In most cases, the system of loop equations \eqref{MMLE} yields more loop variables than independent loop equations. As suggested by Anderson et al. \cite{Anderson:2016rcw}, the positivity of $\langle  \mathcal{O}^\dagger \mathcal{O}\rangle$ can be utilized to constrain the relevant dynamical quantities. We have significantly advanced this method by incorporating the following enhancements:
\begin{enumerate}
    \item We take into account the Back-track loop equations, which will be discussed in more detail later.
    \item The reflection positivity of the lattice system is considered.
    \item Lattice symmetry is utilized to reduce the positivity conditions.
    \item We employ the large $N$ relaxation method proposed by the previous results \cite{Kazakov:2021lel}.
\end{enumerate}

These improvements have allowed us to derive an impressive numerical bound on the one plaquette average $u_P=\frac{1}{N_c}\langle\mathrm{tr} U_P\rangle$. As shown in Fig.~\ref{fig:plaquette4D}, the bootstrap bounds for \(u_P\) for \(L_{\mathrm{max}}=8,12,16\) demonstrate rapid refinement as the cutoffs increase. The upper bound effectively encapsulates the physically significant Wilson loop phase and reliably replicates the 3-loop Perturbation Theory over an extensive coupling range, even exceeding the phase transition point. A comparison with Monte Carlo data, however, indicates room for improvement, especially in the \((2.4, \,2.8)\) interval where the data diverges from the Perturbation Theory. This interval, sourced from \cite{Athenodorou:2021qvs}, was used to compute masses and the string tension. A notable enhancement is anticipated upon reaching \(L_{\mathrm{max}}=20 \text{ or } 24\), although this will necessitate substantial computational resources.

\subsection{Conformal bootstrap}

The author's scholarly pursuits during the PhD career also extend to the conformal bootstrap. 

Recent studies have underscored the stringent constraints CFT data, defining local operator correlators, must adhere to \cite{Poland:2018epd}. These restrictions not only bound the CFT theory space but also intriguingly position salient theories at the edge of the allowable domain \cite{El-Showk:2012cjh}. The conformal bootstrap community has so far investigated this space, mainly probing two directions of constraints on a specific set of four-point correlators. First, we maximize the gap in scaling dimensions of the first operator in the Operator Product Expansion (OPE) \cite{Rattazzi:2008pe}. Second, we bound the OPE coefficient of specific operators within these correlators \cite{Rattazzi:2010gj}. These bounds are established using numerical or analytical functionals, offering dual perspectives on the CFT landscape. This work proposed new directions by bounding CFT correlator values, a natural yet unexplored problem. This approach, similar to the method of minimizing or maximizing $S$-matrix values, is expected to reveal uncharted territories in the CFT space. The $S$-matrix problem, known to yield bounds saturable by exciting theories, is closely tied to (limits of) CFT correlators \cite{Paulos:2016fap}, suggesting equal potential for the corresponding CFT problem. 

\begin{figure}[ht]
  \centering
	\includegraphics[width=15cm]{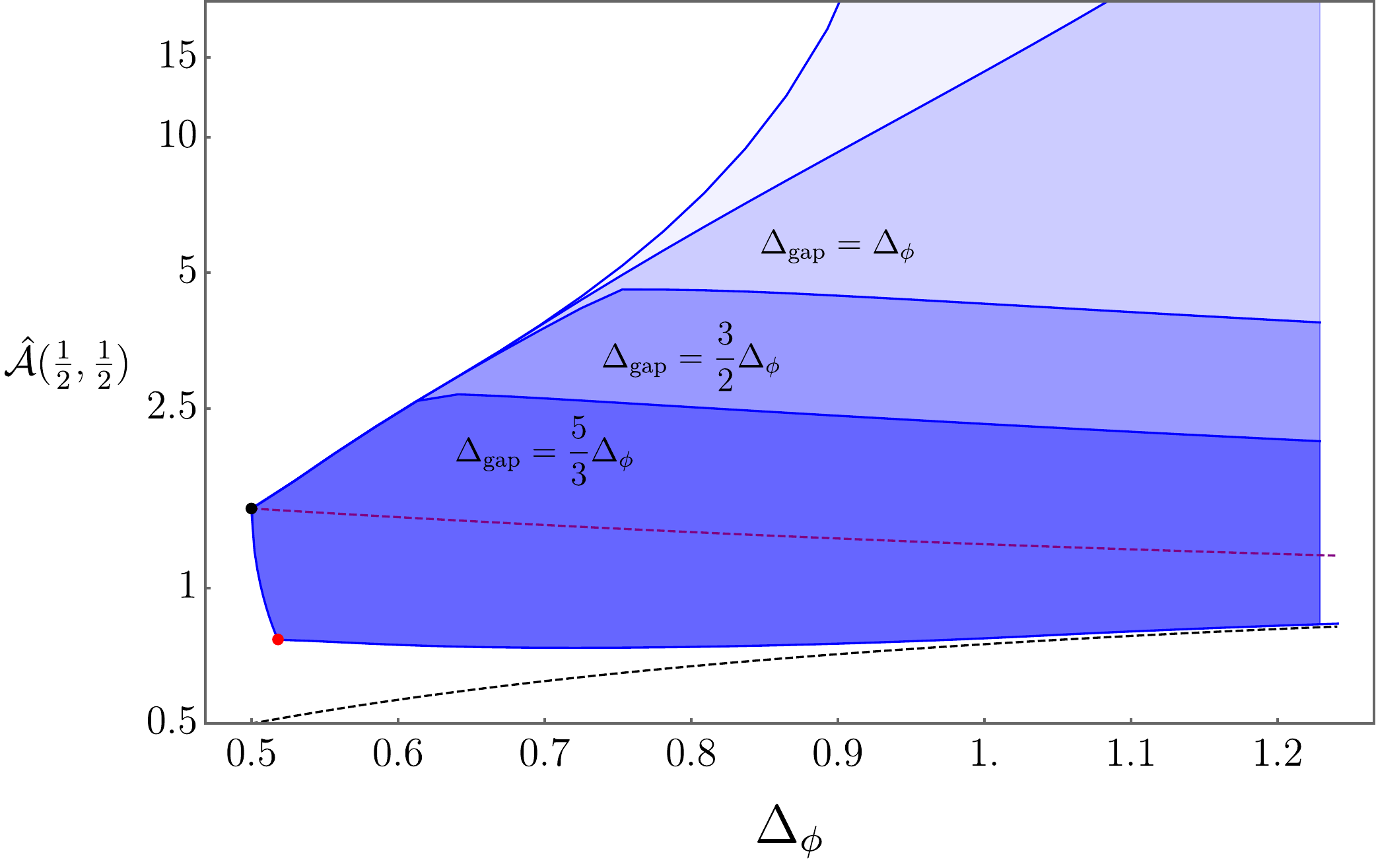}
\caption{\label{allowed}Bounds on 3d CFT correlator values, with $\hat{\mathcal A}(z,\bar z)=(z\bar z)^{\Delta_\phi}\mathcal G(z,\bar z)-1$. The shaded region represents the values such a correlator may take at the crossing symmetric point $z=\bar{z}=\frac 12$. The dashed lines inside the allowed region correspond to maximal allowed values assuming a gap. The upper bound goes to $\infty$ close to $\Delta_\phi=1$. The dashed line outside the allowed region is the 1d generalized free fermion correlator, which provides a (suboptimal) lower bound. The dashed line inside the allowed region is the generalized free boson correlator, which provides an upper bound for $\Delta_{\mathrm{gap}}=2\Delta_\phi$. The black and red dots are the free theory and 3d Ising values respectively.}
\end{figure}

Our numerical findings, captured in Figures \ref{allowed}, depict the four-point correlator's allowed range at the crossing symmetric point $z=\bar z=\frac 12$ as a function of $\Df$, and bounds along $z=\bar z$ for varied $\Df$ values, respectively.

Figure \ref{allowed} reveals an upper correlator bound for some $\Df$ range, irrespective of spectrum assumptions. Imposing gaps in the scalar sector refines this bound, with the generalized free boson correlator closely saturating the upper bound for a $2\Df$ gap, aligning with the exact bound in~\cite{Paulos:2020zxx}\footnote{The argument draws upon the analytic functional, a specific theoretical construct in conformal field theory with an extensive historical foundation~\cite{Afkhami-Jeddi:2020hde, Carmi:2019cub, Caron-Huot:2020adz, Caron-Huot:2022sdy, Dey:2016mcs, Dey:2017fab, El-Showk:2012vjm, El-Showk:2016mxr, Ferrero:2019luz, Ghosh:2021ruh, Ghosh:2023lwe, Ghosh:2023onl, Giombi:2020xah, Gopakumar:2016cpb, Gopakumar:2016wkt, Gopakumar:2021dvg, Hartman:2019pcd, Hartman:2022zik, Kaviraj:2018tfd, Kaviraj:2021cvq, Li:2023whn, Mazac:2016qev, Mazac:2018biw, Mazac:2018mdx, Mazac:2018ycv, Mazac:2019shk, Paulos:2019fkw, Paulos:2019gtx, Paulos:2020zxx, Qiao:2017lkv, Trinh:2021mll}.}. For $\Df=1$, the bound diverges due to the unitary crossing equation solution devoid of identity for $\Df\geq d-2$. This leads to no bound without a spectrum gap assumption, requiring the gap $\Delta_g > \Df/2$.

At $\Df=1/2$, the free theory point, the upper and lower bounds converge with the free theory value, as expected for the single CFT correlator with that dimension. The lower bound is also congruent with the exact bound determined by the 1d generalized free fermion solution, albeit stronger due to the 3d crossing equation constraints.

The lower bound displays a pronounced kink at $\Df\sim 0.518149$, the spin field dimension in the 3d Ising CFT. This reinforces our argument linking correlator minimization with gap maximization, as the latter leads to the 3d Ising model at the precise $\Df$ value:
\begin{equation}
    \Delta_g^{\mbox{\tiny min}}\sim \Delta_{\mbox{\tiny gapmax}}
\end{equation}
\chapter[Matrix Bootstrap and Lattice Bootstrap]{Matrix Bootstrap and Lattice Bootstrap}\label{chp2_chap}
\chaptermark{Matrix Bootstrap and Lattice Bootstrap}


\section{Matrix model and Lattice gauge theory}

Matrix models, primarily renowned for their role in the study of quantum gravity and string theory, serve as a powerful tool for the analytic investigation of numerous complex systems. Large $N$ lattice Yang-Mills theory, on the other hand, is instrumental in our understanding of non-perturbative aspects of quantum chromodynamics (QCD).

In this section, we provide an introductory overview of matrix models and large $N$ lattice Yang-Mills theory, restricting our focus to aspects pertinent to our ensuing bootstrap problem formulation.

\subsection{Matrix Models and the Loop Equations}

Matrix models, including the matrix integrals, are of considerable interest in a multitude of physical and mathematical domains such as multi-component quantum field theory~\cite{tHooft:1973alw} (see \cite{Migdal:1983qrz} for the review), two-dimensional quantum gravity and string theory~\cite{David:1984tx, Kazakov:1985ea, Kazakov:1985ds, Kazakov:1987qg}, mesoscopic physics~\cite{PhysRevLett.52.1}, algebraic geometry~\cite{Dijkgraaf:2002fc, Dijkgraaf:2002pp, Eynard:2007kz, Kontsevich:1992ti}, and number theory~\cite{montgomery1973pair}. A subcategory of matrix models, known as random matrices, have been substantially studied due to their propensity for capturing the statistical behavior of intricate quantum systems.

We focus our attention on one of the principal categories of random matrix models, namely the Hermitian matrix models. A typical $N \times N$ Hermitian one-matrix model can be defined via its probability measure, often represented in the form of a partition function $Z_N$:
\begin{equation}\label{MMint}
    Z_{N}=\int d^{N^2}M\,\e^{-N\tr { V}(M)}
\end{equation}
where $d^{N^2}M=\prod_{i,j 1}^{N}dM_{ij}$ serves as the invariant Hermitian measure. The potential is customarily chosen to be a polynomial:
\begin{equation}\label{pot}
   V(x)=\sum_{k=2}^{d+1} \frac{g_k}{k} M^k.
\end{equation}
The significant "physical observable" in this context is the \(k\)-th moment:
\begin{equation}\label{moment}
    \mathcal{W}_k=\langle \mathrm{Tr }M^k\rangle=\int \frac{d^{N^2}M}{Z_{N}}\,\frac{1}{N}\tr M^k\e^{-N\tr { V}(M)}.
\end{equation}

This model, in the planar limit, is solvable for arbitrary polynomial potentials~\cite{Brezin:1977sv}. Several methods, including direct recursion relations for planar graphs, orthogonal polynomials, the saddle point approximation for the eigenvalue distribution, and loop equations, have been developed to that effect.

In the context of our bootstrap problem, the loop equations derived using the Schwinger-Dyson method assume critical importance. Utilizing the normalized trace $\mathrm{Tr}=\frac{1}{N}\tr$, we express the loop equation in terms of the moments:
\begin{equation}\label{loopN}
    \langle \mathrm{Tr}V^\prime (M)M^k\rangle=\sum_{l=0}^{k-1} \langle \mathrm{Tr}M^l  \mathrm{Tr}M^{k-l-1}\rangle.
\end{equation}
In the \(N\rightarrow \infty\) limit, we employ the factorization property:
\begin{equation}\label{factorN}
    \langle \mathrm{Tr}M^l \mathrm{Tr}M^{m}\rangle= \langle \mathrm{Tr}M^l\rangle \langle \mathrm{Tr}M^{m}\rangle+\mathcal{O}(1/N^2).
\end{equation}
The loop equation then simplifies to:
\begin{equation}\label{loop}
    \sum_{j=1}^{d} g_j\mathcal{\,\,W}_{k+j}=\sum_{l=0}^{k-1} \mathcal{W}_{l} \mathcal{\,\,W}_{k-l+1}.
\end{equation}

While these loop equations constitute a set of coupled nonlinear equations and are generally complex, they simplify in the large $N$ limit and are amenable to solving using techniques such as orthogonal polynomials or topological recursion. The solutions deliver expectation values of the traces of powers of $M$, which encapsulate crucial information regarding physical quantities in the theory, including the free energy and correlators.

In summary, despite the inherent complexity of loop equations, their solutions afford valuable physical insights, underlining the power of matrix models as analytical tools in the study of complex quantum systems.

\subsection{Large N Lattice Yang-Mills Theory}

Quantum Chromodynamics (QCD), the theory of strong interactions, is an example of a non-Abelian gauge theory (a Yang-Mills theory) based on the $SU(3)$ gauge group. Generalizing this to an $SU(N)$ gauge group allows us to study the theory in the large $N$ limit, which provides both conceptual and computational simplifications. However, direct analytical treatment of large $N$ QCD, or any non-Abelian gauge theory, is challenging due to the strong coupling at low energies. A common approach to overcome this issue is to put the theory on a lattice, leading to the so-called Lattice Gauge Theory~\cite{,Wilson:1974sk}\footnote{For a review of other realization and generalization of the lattice Yang-Mills theory, the reader could refer to~\cite{Cao:2023uqm}.}: 

\begin{equation}\label{eq: action}
 S=-\frac{N_c}{\lambda}\sum_{P} {\rm Re}\,\tr U_P\,
\end{equation}

Lattice Gauge Theory provides a non-perturbative formulation of Yang-Mills theory where spacetime is discretized to a lattice of points. On this lattice, the gauge fields are defined on the links connecting the points and the fermion fields at the sites.

In the large $N$ limit, Lattice Gauge Theory simplifies significantly. This limit suppresses quantum fluctuations, allowing a semi-classical treatment\cite{Gopakumar:1994iq}. Moreover, it provides a systematic $1/N$ expansion where each order corresponds to a specific class of Feynman diagrams (planar diagrams at leading order).

In practice, Lattice Gauge Theory, especially in the large $N$ limit, is often studied using numerical simulations. The path integral is evaluated stochastically using techniques such as Monte Carlo sampling\cite{Creutz:1980zw}. Large $N$ reduces the sign problem, making these simulations more feasible.

The large $N$ limit of Lattice Gauge Theories also has a deep connection to Matrix Models. This limit simplifies the Lattice Gauge Theory to a Matrix Model, where the degrees of freedom are matrices rather than fields\cite{Bhanot:1982sh, Eguchi:1982nm, Gonzalez-Arroyo:2010omx}. This opens the way to study non-perturbative aspects of gauge theories using the powerful methods developed in the context of Matrix Models.

In conclusion, the large $N$ Lattice Yang-Mills theory provides a unique window to non-perturbative aspects of gauge theories. It is a broad and active field of research, with connections to areas as diverse as String Theory and Statistical Mechanics.

\section{Bootstrapping a toy model}

To be concrete, here we briefly illustrate "Bootstrap by equations of motion" by a toy model: 0-dimensional $\varphi^4$ model. Consider the model defined by the following partition function:
\begin{equation}
    Z=\int_{-\infty}^{\infty} \exp (-\frac{x^2}{2}-g\frac{x^4}{4})\mathrm{d}x,\,\ g>0,
\end{equation}
and our goal here is solving the observables:
\begin{equation}
    \mathcal{W}_k=\frac{1}{Z}\int_{-\infty}^{\infty} x^k\exp (-\frac{x^2}{2}-g\frac{x^4}{4})\mathrm{d}x.
\end{equation}
Due to the simplicity of the model, we have a lot of available method to solve the model. For a closed-form solution, we have:

\begin{equation}\label{eq:close}
    \mathcal{W}_2=\frac{\pi  \left(-I_{-\frac{1}{4}}\left(\frac{1}{8 g}\right)+(4 g+1)
   I_{\frac{1}{4}}\left(\frac{1}{8
   g}\right)-I_{\frac{3}{4}}\left(\frac{1}{8
   g}\right)+I_{\frac{5}{4}}\left(\frac{1}{8 g}\right)\right)}{2 \sqrt{2}
   g K_{\frac{1}{4}}\left(\frac{1}{8 g}\right)}
\end{equation}

\begin{figure}[h]
    \begin{flushright}
    \includegraphics[width=0.8\textwidth]{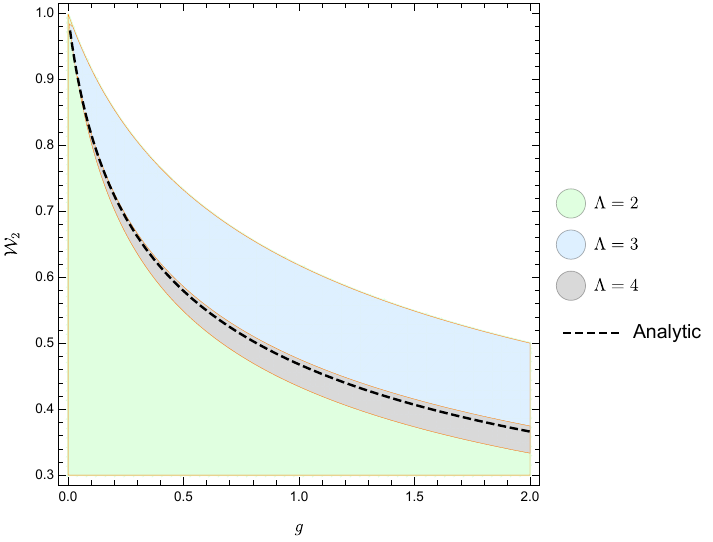}  
    \caption{The shrinking of the allowed region for the toy model.}
    \label{fig: toy}
    \end{flushright}
\end{figure}

The bootstrap method starts by considering the Dyson-Schwinger equations (or by integration by part):
\begin{equation}
    (k+1)\mathcal{W}_k=\mathcal{W}_{k+2}+g\mathcal{W}_{k+4}
\end{equation}
and global symmetry:
\begin{equation}
    \mathcal{W}_k=0,\, \mathrm{for}\, \mathrm{odd}\,k
\end{equation}
to solve all the higher moments in terms of \(\mathcal{W}_2\). To formulate the problem as  an optimization problem, a crucial step is considering the expectations of square of polynomials are always positive semi-definite:
\begin{equation}
    \frac{1}{Z}\int_{-\infty}^{\infty} (\sum \alpha_i x^i)^2\exp (-\frac{x^2}{2}-g\frac{x^4}{4})\geq 0,\, \forall \alpha
\end{equation}
This is a quadratic form in \(\alpha\), so its positivity is equivalent to:
\begin{equation}
    \mathbb{W}=
    \begin{pmatrix} 
    \mathcal{W}_0 & \mathcal{W}_1 & \mathcal{W}_2& \dots \\
    \mathcal{W}_1 & \mathcal{W}_2 & \mathcal{W}_3& \dots \\
    \mathcal{W}_2 & \mathcal{W}_3 & \mathcal{W}_4& \dots \\
    \vdots &\vdots&\vdots &\ddots &  
    \end{pmatrix}\succeq 0
\end{equation}
We can solve the Semi-Definite Programming(SDP) maximizing or minimizing \(\mathcal{W}_2\) constrained by a truncation of the matrix $\mathbb{W}$:
    \begin{align}
        \min \mathrm{or} &\max \,\mathcal{W}_2\\
         &\mathbb{W}_{\Lambda}\succeq 0
    \end{align}
Here \(\mathbb{W}_{\Lambda}\) is the top \((\Lambda+1)\times (\Lambda+1)\) sub-matrix of \(\mathbb{W}\).

For \(g=1,\, \Lambda=10\), we can get the numerical bootstrap result (5.5 digits):
\begin{equation}
    0.4679137 \leq \mathcal{W}_2=0.4679199170 \leq 0.4679214
\end{equation}
The value in the middle is from the closed-form solution Eq~\ref{eq:close} (exact value). The l.h.s and r.h.s is the bootstrap result maximizing and minimizing $\mathcal{W}_2$.

This simple toy model (cannot be even simpler) is actually the prototype of the bootstrap implementations fall in the category of ``Bootstrap by Equations of Motion''. We see that it could be fairly general, i.e. this method can potentially work for any system defined by a measure satisfy certain positivity conditions. 

\newpage
\thispagestyle{empty} 
\addcontentsline{toc}{section}{Paper I}
\begin{center}
\vspace*{\fill}
\textbf{\Huge Paper I}
\vspace*{\fill}
\end{center}
\newpage

\includepdf[pages=-]{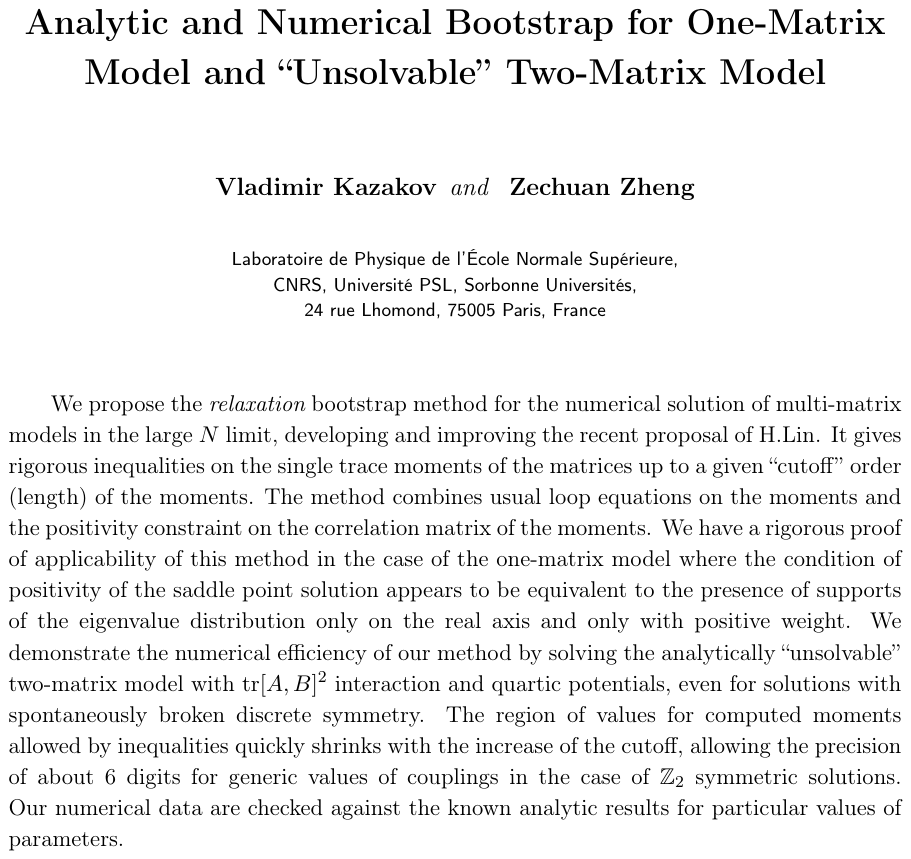}

\newpage
\thispagestyle{empty} 
\addcontentsline{toc}{section}{Paper II}
\begin{center}
\vspace*{\fill}
\textbf{\Huge Paper II}
\vspace*{\fill}
\end{center}
\newpage

\includepdf[pages=-]{ch2/fig/YMBootstrap.pdf}

\chapter[Conformal Bootstrap]{Conformal Bootstrap}\label{chp2_chap}
\chaptermark{Conformal Bootstrap}


\section{Conformal Bootstrap for conformal correlators}

The conformal bootstrap is a potent tool for investigating and constraining the properties of Conformal Field Theories (CFTs). The origins of this method can be traced back to the 1970s, stemming from the works of Polyakov, Ferrara, Gatto, Grillo, and others \cite{Ferrara:1973yt, 1974JETP...39...10P}, who utilized the properties of operator product expansions (OPEs) and conformal symmetry to derive constraints on CFTs.

\subsection{Basic Idea}
The fundamental idea of the conformal bootstrap is the use of symmetries and consistency conditions to impose constraints on, or even solve, CFTs. The linchpin of this method is the crossing symmetry equation, derived from the associativity of the OPE and the invariance under the interchange of operators in the correlation function.

\subsection{Core Premise}
The essential premise \cite{Rattazzi:2008pe} focuses on analyzing the constraints derived from the crossing symmetric equation:

\begin{equation}
\sum_{\Delta,\ell} a_{\Delta,\ell} F_{\Delta,\ell}(z,\bar{z})=0,
\end{equation}

Here, the crossing vector $F_{\Delta,\ell}(z,\bar{z})$ is determined by the crossing symmetry:

\begin{equation}
F_{\Delta,\ell}(z,\bar{z})=\frac{G_{\Delta,\ell}(z,\bar{z})}{(z\bar{z})^{\Delta_\phi}}-\frac{G_{\Delta,\ell}(1-z,1-\bar{z})}{\left((1-z)(1-\bar{z})\right)^{\Delta_\phi}}
\end{equation}

To facilitate analysis, this equation is typically discretized at the crossing-symmetric point $z=\bar{z}=\frac{1}{2}$. This process employs a variety of inventive techniques to compute these derivatives efficiently \cite{Kravchuk:2016qvl, Erramilli:2019njx, Kravchuk:2017dzd, Karateev:2018oml, Erramilli:2020rlr}. Post discretization, polynomial approximation transforms the equation into a standard Semidefinite Program (SDP) constraint condition. The \textit{SDPB} solver \cite{Simmons-Duffin:2015qma} provides the solution to the corresponding SDP with arbitrary precision. With these constraints, the `navigitor' technique \cite{Reehorst:2021ykw} explores physically permissible theories within the CFT parameter space.

\subsection{Challenges in Correlator Bounding}
However, this established paradigm faces complications when applied to bound the correlator. When formulating the objective function to bound the correlator as follows:

\begin{equation}
\text{max/min: } \sum_{\Delta,L} a_{\Delta,L} G_{\Delta,L}(z, \bar{z})
\end{equation}

and $(z, \bar{z})$ is not at the crossing symmetric point, the objective function and the crossing symmetry equation exhibit different exponential decay with respect to the $\Delta$. Consequently, no valid rational approximation exists for the bootstrap formulation. To address this, the work presented here employed the continuous primal simplex algorithm \cite{El-Showk:2014dwa}, implemented in \textit{JuliBoots}\cite{2014arXiv1412.4127P}.

\newpage
\thispagestyle{empty} 
\addcontentsline{toc}{section}{Paper III}
\begin{center}
\vspace*{\fill}
\textbf{\Huge Paper III}
\vspace*{\fill}
\end{center}
\newpage

\includepdf[pages=-]{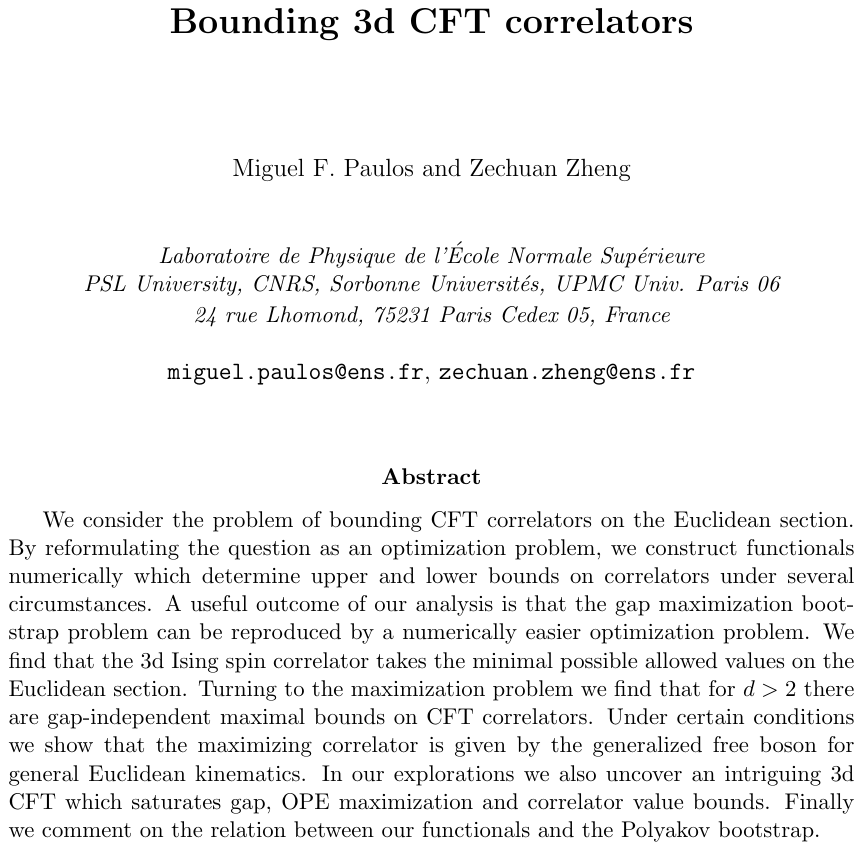}

\backmatter
\addchapnonumber{Conclusion and Prospects}


This dissertation has thoroughly explored the application of the bootstrap method to matrix models, Yang-Mills theory, and conformal field theory. We have underscored the potency and versatility of the bootstrap method as an optimization-based approach to problem-solving in the contemporary physical landscape, often yielding precise numerical bounds and deep theoretical insights.

We proposed an enhanced relaxation bootstrap method to numerically solve multi-matrix models in the large N limit. We showed that this method offers robust inequalities on the single trace moments of the matrices up to a specified “cutoff” order of the moments. Our demonstration of the method's numerical efficiency is made evident through the successful resolution of the analytically “unsolvable” two-matrix model with tr[A, B]² interaction and quartic potentials.

The extension of our study to lattice Yang-Mills theory in dimensions 2, 3, and 4 using the numerical bootstrap method has yielded promising outcomes. The combination of loop equations with a cutoff on the maximal length of loops and positivity conditions on specific matrices of Wilson loop averages has demonstrated that this approach could present a tangible alternative to the predominant Monte Carlo approach.

Our investigation into bounding CFT correlators in the Euclidean section elucidates a remarkable connection between optimization problems and the determination of upper and lower bounds on correlators. Our work suggests that the 3d Ising correlator occupies the minimal possible allowed values on the Euclidean section. We have also unveiled a peculiar 3d CFT that saturates gap, OPE maximization, and correlator value bounds.

In conclusion, this thesis has deepened our understanding of the power and versatility of the bootstrap method in dealing with complex problems in contemporary physics. The results obtained in matrix models, Yang-Mills theory, and conformal field theory demonstrate the considerable promise of this approach, and we anticipate that further exploration will continue to yield valuable insights into the fundamental nature of these systems. The journey of discovery using the bootstrap method is far from complete, and the results of this dissertation provide a strong foundation for future explorations and advancements.

Despite the promising direction, the bootstrap method is still in its early stages and is open to significant improvements. The following improvements are proposed to achieve the ambitious goals outlined:

$\bullet{}$ The establishment of a general method to bootstrap the long-range/asymptotic behavior of lattice systems. The existing lattice bootstrap results \cite{Kazakov:2022xuh, Cho:2022lcj} primarily focus on the correlators localized in small regions. A comprehensive method to bootstrap long-range behavior of correlators is required, which will encapsulate the information in the IR of the lattice theories.

$\bullet{}$ Parallelization. The current algorithm, albeit accurate, is essentially sequential. Notably, most algorithm components can be readily parallelized. Employing MPI or CUDA libraries for complete parallelization of the algorithm would markedly enhance computational efficiency.

The bootstrap method studied in this thesis, in its nascent stages, has already demonstrated its capacity to yield highly accurate numerical results with $100\%$ error control. This thesis asserts the high potentiality of the bootstrap method in numerous areas of physics. This document delineates several anticipated applications and goals for this method.

\section{Lattice Field Theory}
Lattice Quantum Chromodynamics (QCD) is an instrumental framework in comprehending the non-perturbative aspects of QCD. Despite its successes, the standard numerical method for Lattice QCD - the Monte Carlo (MC) method, faces inherent limitations. These include statistical errors, finite lattice size, computational burdens for dynamic quark inclusion, challenges in addressing finite baryon density, and real-time dynamics.

Our recent result \cite{Kazakov:2022xuh} suggests the bootstrap approach could provide a viable alternative or even pose competition for the MC method. An immediate and significant research direction is to generalize large N bootstrap settings to finite color numbers, with a particular interest in $N_c=3$. Potential quantities like quark condensation $\langle \bar{\psi}(0) \psi(0)\rangle$ \cite{Cvetic:1997eb} are already available for comparison in the pure gauge sector.

We believe another bootstrap candidate is the glueball spectrum \cite{Athenodorou:2021qvs}. The bootstrap method can offer rigorous results for the asymptotic region, surpassing traditional numerical methods without necessitating extrapolation, as demonstrated in the large N limit \cite{Kazakov:2021lel}. We posit that the bootstrap method can also compute glueball masses, which are asymptotic quantities in the large volume limit.

Furthermore, the bootstrap method can be applied to finite temperature lattice QCD, with an aim to establish rigorous bounds on quantities like the Polyakov loop, which could shed light on the confinement phenomenon in QCD.

In addition, there is potential for a bootstrap formulation of the (renormalizable) quantum field theory. While lattice field theory is one of the potential regularizations of UV physics, the bootstrap method's efficiency is closely tied with symmetries. Consequently, lattice regularization, which disrupts the original symmetries of the quantum field theory, may not be optimal. It would be of significant interest to generalize this bootstrap method to alternative regularizations that conserve the original symmetries of quantum field theory.

\section{Quantum Systems on the Lattice}
For certain problems in condensed matter physics, there is an acute need for reliable numerical tools to understand the model's ground state characterizations and phase diagrams.

Consider the Hubbard model in $d\geq 2$. This model is believed to exhibit fascinating physical phenomena like superconductivity \cite{2021arXiv210312097A}, but its phase diagram remains elusive.

A similar case is the $2d$ quantum Heisenberg model on the kagome lattice. Previous studies suggested that its ground state could be a gapped $\mathbb{Z}_2$ spin liquid \cite{2011Sci...332.1173Y}. However, recent numerical evidence supports the ground state realization as a gapless $U(1)$ Dirac spin liquid \cite{2017PhRvX...7c1020H}. An efficient numerical approach with rigorous error control will unquestionably deepen our understanding of these models.

\addchapnonumber{Résumé en Français}\label{resumeF}

\section{Contexte}

Dans le domaine de la théorie quantique des champs (TQC) moderne, l'importance de la symétrie est indiscutable. Notre compréhension progressive de la TQC a toujours été associée aux connaissances acquises grâce aux symétries, qu'elles soient supposées (comme la symétrie de Poincaré et le pouvoir contraignant des anomalies) ou brisées spontanément.

\begin{figure}[h!]
\centering
\includegraphics[scale=.25]{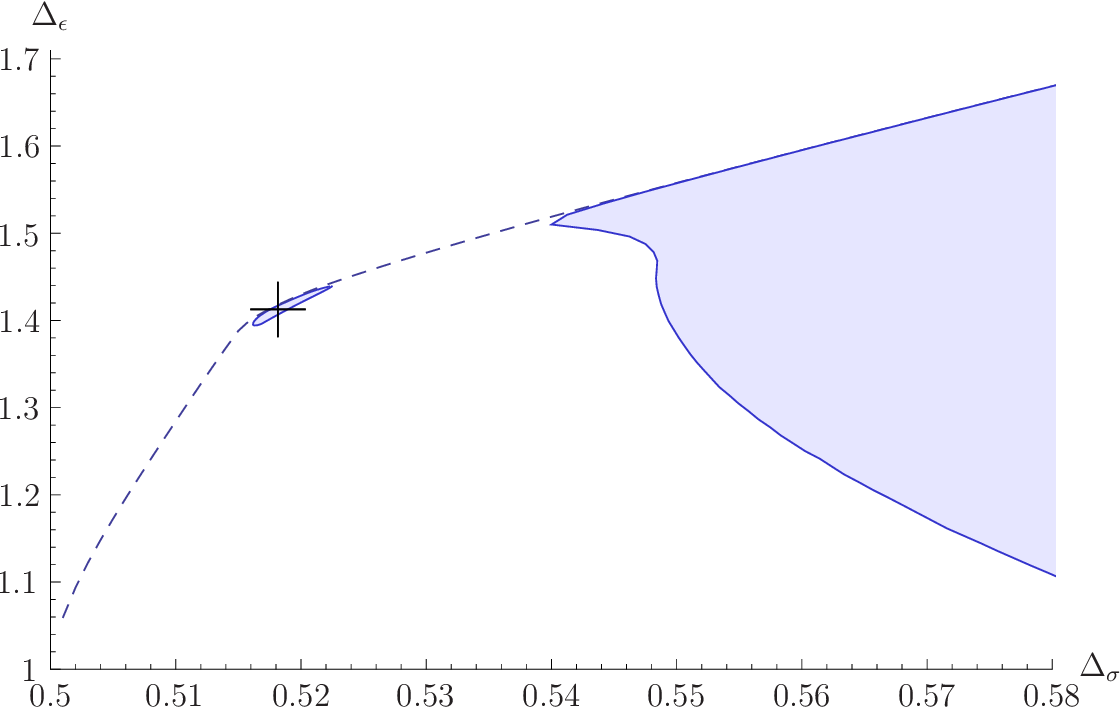}
\caption{La région représentée correspond à l'espace des paramètres autorisé par le modèle Ising en $3d$. Elle est définie par les contraintes de trois corr\'elateurs à quatre points distincts, en supposant qu'il n'y a qu'un seul opérateur pertinent $\mathbb{Z}_2$-impair présent. Le graphique a été adapté du travail de Kos et al. \cite{Kos:2014bka}.}
\label{fig:ising}
\end{figure}

La philosophie du bootstrap en physique incarne une méthodologie qui vise à exploiter la puissance potentielle de la symétrie à son maximum. L'approche cherche à résoudre les problèmes principalement en utilisant l'hypothèse de symétrie, avec d'autres contraintes génériques comme l'unitarité et la localité. Cette méthode évite de s'appuyer sur les détails microscopiques du problème en question. Bien que cette perspective ait été initialement proposée dans les années 1960 dans le contexte de la S-matrice \cite{Chew:110879}, elle a été largement éclipsée par les succès ultérieurs en Chromodynamique Quantique (CDQ).

Le premier résultat significatif suivant la philosophie du bootstrap n'a pas vu le jour avant les années 1980, lorsque Belavin, Polyakov et Zamolodchikov (BPZ) ont utilisé le "bootstrap conforme" pour résoudre des modèles minimaux avec $c<1$ \cite{belavin1984infinite}. Leur solution reposait fortement sur les symétries infinies présentes dans la théorie des champs conformes en $2d$, une caractéristique absente dans d'autres dimensions.

Le 21ème siècle a connu une augmentation sans précédent du développement de la théorie de l'optimisation mathématique, largement stimulée par l'expansion rapide de l'industrie de l'apprentissage automatique, qui a connu une percée significative en 2006 \cite{10.1162/neco.2006.18.7.1527}. À peu près à la même période, un travail fondateur a émergé, utilisant la théorie de l'optimisation pour limiter la dynamique de la théorie des champs conformes en $4d$ (TCC) \cite{Rattazzi:2008pe}. Cette méthode, connue sous le nom de "méthode bootstrap", a suscité un intérêt renouvelé pour l'application de la théorie de l'optimisation pour aborder les problèmes en physique théorique \cite{Ferrara:1973yt, 1974JETP...39...10P}. Malgré sa nature numérique, la méthode bootstrap s'est avérée efficace pour générer des limites rigoureuses pour les quantités dynamiques. L'approche a connu un succès particulièrement remarquable dans le contexte de la TCC \footnote{Des revues complètes de ces avancées peuvent être trouvées dans \cite{Poland:2018epd, Bissi:2022mrs}. Voir aussi \cite{Poland:2022qrs, Hartman:2022zik} pour des mises à jour plus récentes.}, démontrant que des résultats numériques très précis peuvent être obtenus uniquement à partir des axiomes de la TCC, de l'unitarité et des symétries globales.

L'invariance conforme et l'expansion du produit d'opérateurs ont conduit à l'équation dite de bootstrap :
\begin{equation}\label{boot}
\sum_{\Delta} a_{\Delta}^2 F_{\Delta}\left(z,\bar{z}\right)=0,\quad (z,\bar{z})\in (\mathbb{C}\backslash (-\infty,0)\cup(1,\infty))^2
\end{equation}
Ici, l'invariance conforme nous aide à déterminer $F_{\Delta}[z,\bar{z}]$ (soit analytiquement soit numériquement), et l'unitarité nous donne la réalité et la région potentielle $\Delta$ dans l'Eq. (\ref{boot}).

La situation mentionnée représente une généralisation infinie de la programmation linéaire. À partir de cela, nous pouvons obtenir des limites pour des quantités dynamiques intrigantes comme l'écart de dimension au-dessus de l'opérateur d'identité. Par exemple, considérez une fonctionnelle $\Lambda$ agissant sur $F_{\Delta}$, où $\Lambda(F_{\Delta})\geq 0$ pour $\Delta\geq \Delta_{\mathrm{gap}}$ et $\Lambda(F_{0})> 0$. L'existence d'une telle fonctionnelle, couplée à l'Eq. (\ref{boot}), implique qu'il doit y avoir au moins un opérateur dans la région $0<\Delta<\Delta_{\mathrm{gap}}$. En recherchant $\Lambda$ sur tous les $\Delta_{\mathrm{gap}}$ possibles, nous pouvons obtenir une limite optimale pour l'écart de dimension.

De nombreuses études ultérieures utilisant l'approche bootstrap ont produit des résultats convaincants, montrant l'efficacité de cette méthodologie. L'un des réalisations les plus significatives dans cette veine a été la détermination des dimensions d'opérateur dans le modèle Ising Critique 3D avec une précision sans précédent \cite{El-Showk:2012cjh, El-Showk:2014dwa, Kos:2014bka}. Comme représenté sur la Fig. \ref{fig:ising}, le modèle Ising critique en $3d$, tel que déterminé par le bootstrap conforme, réside dans une minuscule île de l'espace des paramètres.

\begin{figure}[h]
    \centering
    \includegraphics[width=0.45\textwidth]{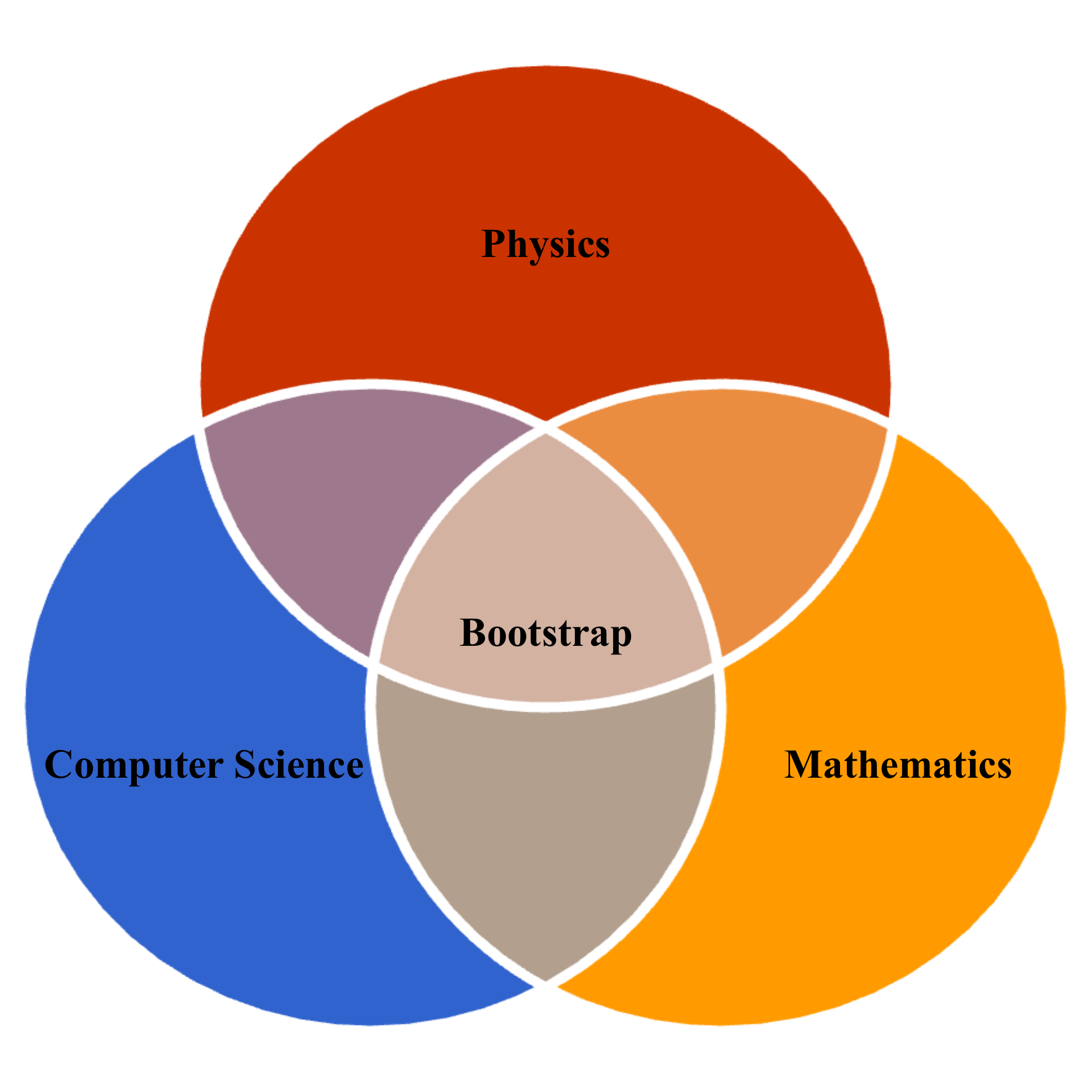}
    \caption{La méthode bootstrap.}
    \label{fig: relationship}
\end{figure}

L'exécution réussie de la méthode bootstrap, illustrée sur la Figure~\ref{fig: relationship}, témoigne de l'interaction essentielle entre la physique, les mathématiques et l'informatique :

\begin{itemize}
\item Sur le front de la physique, une compréhension complète des principes sous-jacents est essentielle pour concevoir des contraintes efficaces. Des exemples notables de ceci incluent l'application de l'expansion du bloc conforme\cite{Dolan:2003hv} du corr\'elateur à quatre points en théorie des champs conformes, et l'utilisation des équations de boucle de Makeenko-Migdal\cite{Makeenko:1979pb} en théorie de jauge comme contraintes non triviales dans les calculs de bootstrap.

\item Les mathématiques fournissent l'ensemble d'outils nécessaires pour interpréter et valider les contraintes associées au modèle physique. Par exemple, une solide analyse mathématique est requise pour vérifier la véracité numérique du bootstrap conforme \cite{Pappadopulo:2012jk}.

\item Dans le domaine de l'informatique, des techniques robustes sont essentielles pour la mise en œuvre pratique de la méthode bootstrap. Une maîtrise de la programmation est nécessaire pour le développement de solveurs d'optimisation spécialisés, tels que \textit{SDPB} \cite{Simmons-Duffin:2015qma, Reehorst:2021ykw, Liu:2023elz}, \textit{JuliBoots} \cite{2014arXiv1412.4127P}, \textit{PyCFTBoot}\cite{Behan:2016dtz} et \textit{FunBoot} \cite{Ghosh:2023onl}, spécifiquement conçus pour résoudre les problèmes de bootstrap. Il est également à noter que d'importants résultats numériques issus de la méthode bootstrap sont généralement générés à l'aide de ressources de supercalcul \cite{Chester:2019ifh}.
\end{itemize}

Malgré les avancées impressionnantes dans le bootstrap conforme, l'application de la méthode bootstrap à des systèmes dépourvus de l'avantage de la symétrie conforme reste un problème difficile. Un domaine potentiel d'exploration est le bootstrap de la S-matrice \cite{Homrich:2019cbt, Kruczenski:2022lot, Paulos:2016but, Paulos:2016fap, Paulos:2017fhb}, où l'auteur s'est lancé dans une enquête préliminaire dans ce domaine pendant ses études de master \cite{Paulos:2018fym}. Cependant, les progrès dans le domaine du bootstrap de la S-matrice ont été largement limités à l'établissement de limites générales sur les paramètres de la théorie. À l'heure actuelle, il n'est pas possible d'intégrer efficacement les informations ou les contraintes dérivées de la théorie ultraviolette (UV) dans le cadre du bootstrap de la S-matrice.

À la lumière des avancées récentes \cite{Anderson:2016rcw, Lin:2020mme, Han:2020bkb}, une nouvelle approche bootstrap est apparue pour étudier les modèles matriciels. Cette méthode est caractérisée non seulement par l'adhésion à des hypothèses générales telles que l'unitarité et les symétries globales, mais aussi par l'intégration des relations entre les observables physiques imposées par les équations du mouvement.\footnote{Pour une étude détaillée sur les convergences mathématiques de cette méthode, les lecteurs sont dirigés vers \cite{2022arXiv221005239G, Kazakov:2021lel, Cho:2023ulr}.} Son applicabilité a été rapidement démontrée dans les théories des champs sur réseau \cite{Anderson:2016rcw, Anderson:2018xuq, Cho:2022lcj, Kazakov:2022xuh}, les modèles matriciels \cite{Han:2020bkb, Jevicki:1982jj, Jevicki:1983wu, Koch:2021yeb, Lin:2020mme, Lin:2023owt, Mathaba:2023non}, les systèmes quantiques\cite{Aikawa:2021eai, Aikawa:2021qbl, Bai:2022yfv, Berenstein:2021dyf, Berenstein:2021loy, Berenstein:2022unr, Berenstein:2022ygg, Berenstein:2023ppj, Bhattacharya:2021btd, Blacker:2022szo, Ding:2023gxu, Du:2021hfw, Eisert:2023hcx, Fawzi:2023ajw, hanQuantumManybodyBootstrap2020, Hastings:2021ygw, Hastings:2022xzx, Hessam:2021byc, Hu:2022keu, Khan:2022uyz, Kull:2022wof, Li:2022prn, Li:2023nip, Morita:2022zuy, Nakayama:2022ahr, Nancarrow:2022wdr, Tavakoli:2023cdt, Tchoumakov:2021mnh, Fan:2023bld, Fan:2023tlh, John:2023him, Li:2023ewe, Zeng:2023jek}, et même les systèmes dynamiques classiques \cite{goluskinBoundingAveragesRigorously2018, goluskinBoundingExtremaGlobal2020, tobascoOptimalBoundsExtremal2018, Cho:2023xxx}.


La première partie de cette thèse discute de notre application de cette nouvelle méthode bootstrap aux modèles matriciels \cite{Kazakov:2021lel} et aux théories de jauge de grande $N$ sur réseau \cite{Kazakov:2022xuh}. Dans notre première publication, nous avons justifié rigoureusement l'approche bootstrap dans le contexte d'un modèle à une matrice et introduit un assouplissement sur les termes quadratiques dans les équations de boucle, atteignant ainsi une précision sans précédent dans un modèle à deux matrices insoluble. Dans notre seconde publication, nous avons étendu l'application de cette méthode bootstrap (avec l'assouplissement pertinent) à la théorie de jauge de grande $N$ sur réseau. De manière remarquable, cette approche a donné d'excellents résultats dans l'estimation (ou plus précisément, la délimitation) de la moyenne des plaquettes de la théorie de jauge sur réseau.

La dernière partie de cette thèse aborde le travail collaboratif de l'auteur avec Miguel Paulos, axé sur la délimitation des fonctions de corrélation via le bootstrap conforme\cite{Paulos:2021jxx}. Étant donné le comportement exponentiel unique de la fonction cible dans ce problème bootstrap spécifique, l'approximation polynomiale traditionnelle utilisée dans \textit{SDPB}\cite{Simmons-Duffin:2015qma} est inadaptée. Nous avons démontré que la solution pour maximiser l'écart sature généralement la minimisation du corr\'elateur. En même temps, la maximisation du corr\'elateur, soumise à des contraintes supplémentaires, reproduit efficacement la théorie du champ moyen.

Dans le dernier chapitre de cette thèse, nous consolidons nos découvertes, offrant un résumé concis et les implications de notre travail. De plus, nous suggérons une variété de directions prometteuses pour la recherche et l'exploration futures.


\section{Principaux résultats}
\subsection{Bootstrap matriciel}

Une grande partie de mon programme de recherche tourne autour du bootstrap matriciel \cite{Kazakov:2021lel}. Cette entreprise comprend deux composants cruciaux : une solide justification de la méthode bootstrap pour le modèle à une matrice en utilisant les résultats du problème du moment de Hamburger, et une avancée notable de cette méthode pour bootstrapper les modèles matriciels de grande $N$ via relaxation convexe.

Plus précisément, le processus de justification, s'appuyant sur les résultats du problème du moment de Hamburger, affirme rigoureusement que la positivité de la matrice de corrélation implique la positivité du résolvant, et vice versa :

\begin{equation}\label{equiv}
    \textit{Positivité de la matrice de corrélation} \Leftrightarrow \textit{Positivité du Résolvant}
\end{equation}

Cette équivalence a été utilisée en conjonction avec les équations de boucle pour catégoriser analytiquement et exhaustivement les solutions du modèle matriciel de grande $N$ spécifié ci-dessous :

\begin{equation}\label{MMint}
    Z_{N}=\int d^{N^2}M\,\mathrm{e}^{-N\mathrm{tr} { V}(M)} , \quad V(x)=\frac{1}{2}\mu x^2 +\frac{1}{4}g x^4
\end{equation}

La seconde moitié de l'article se concentre sur la résolution du modèle suivant :

\begin{equation}\label{2MMcom21}
Z=\lim_{N\rightarrow \infty}\int d^{N^2}A,d^{N^2}B,\mathrm{e}^{-N \mathrm{tr}\left( -h[A,B]^2/2+A^2/2+g A^4/4+B^2/2+g B^4/4\right)}
\end{equation}

Avec nos méthodologies existantes, ce modèle n'a pas de solution analytique. Pour bootstrapper ce modèle, nous avons utilisé une méthode de relaxation comme suit :

\begin{equation}\label{relaxm}
    Q=x x^\mathrm{T}\Rightarrow\mathcal{R}=\begin{pmatrix}
1 & x^{\mathrm{T}}\\
x & Q
\end{pmatrix}\succeq 0.
\end{equation}

\begin{figure}
\centering
\includegraphics[scale=.65]{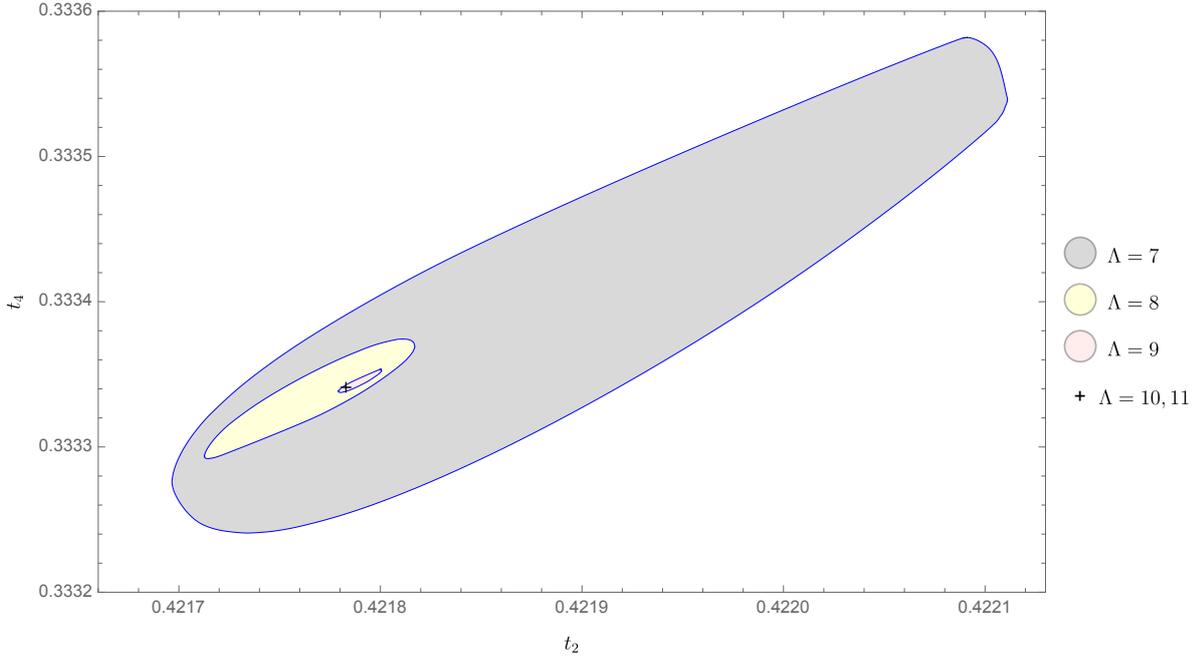}
\caption{La région autorisée de \(t_2-t_4\) du modèle~\eqref{2MMcom21} avec le paramètre $g=1, h=1$ pour le cutoff  \(\Lambda=7,8,9,10,11.\) Nous rappelons la définition de \(\Lambda\) : les opérateurs les plus longs dans la matrice de corrélation et dans les équations de boucle ont la longueur $2\Lambda$.}
\label{fig: covg}
\end{figure}

Dans ce qui précède, $Q$ représente les termes quadratiques dans les équations de boucle, dérivés de la factorisation de grande $N$. Le vecteur $x$ fait référence au vecteur colonne des variables de trace unique. Les résultats obtenus à partir du problème de bootstrap relaxé surpassent considérablement la précision et l'efficacité de la méthode numérique conventionnelle pour les modèles matriciels de grande $N$, la méthode Monte Carlo (MC) \cite{Jha:2021exo}. Pour $g=h=1$, nous avons obtenu un résultat de précision à 6 chiffres :

\begin{equation}
    \begin{cases}
    0.421783612\leq t_2 \leq 0.421784687\\
    0.333341358\leq t_4 \leq 0.333342131
    \end{cases}
\end{equation}

L'illustration de la contraction du domaine admissible en fonction du cutoff bootstrap correspondant est visuellement démontrée dans la Figure~\ref{fig: covg}.
\subsection{Bootstrap de la théorie de Yang-Mills sur réseau}
Un autre projet \cite{Kazakov:2022xuh} a consisté à utiliser cette méthode pour bootstrap la boucle de Wilson à une plaquette moyenne dans la théorie Yang-Mills sur réseau de grande $N$.\footnote{Pour d'autres études de bootstrap de la théorie de jauge ou QCD, le lecteur peut se référer à \cite{Nakayama:2014sba, Albert:2022oes, Albert:2023jtd, Fernandez:2022kzi, Guerrieri:2018uew, He:2023lyy, Caron-Huot:2023tpw, Ma:2023vgc}. Il est largement admis que le Yang-Mills de grande N est équivalent au QCD de grande N en raison de la suppression de la composante fermionique à la limite planaire. Cependant, des preuves récentes suggèrent que cette hypothèse pourrait être plus nuancée~\cite{Cherman:2022eml}.} Les résultats sont encourageants par rapport à la méthode MC, surtout si l'on considère que la méthode MC pour le QCD sur réseau a fait l'objet d'intenses recherches pendant plusieurs décennies.

\begin{figure}[h]
\centering
    \includegraphics[width=.75\textwidth]{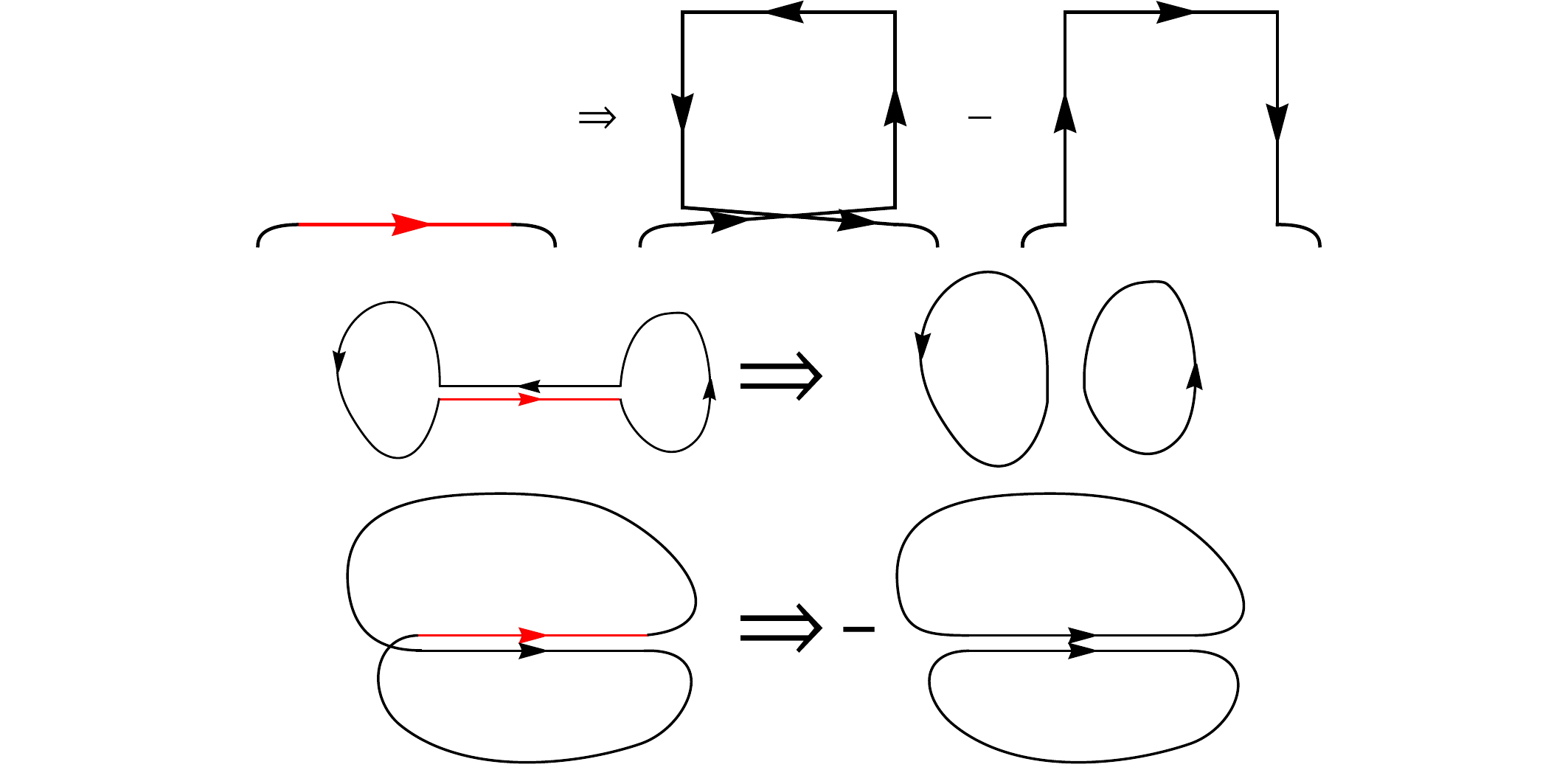}
    \caption{Représentation schématique des LEs : La première ligne montre la variation d'un lien de boucle de Wilson dans le LHS de l'Eq.\eqref{MMLE}. Les 2ème et 3ème lignes montrent la division du contour le long de la ligne variée en deux sous-contours, pour deux orientations différentes des liens coïncidant dans le RHS de l'Eq.\eqref{MMLE}.}
 \label{fig:LE}   
\end{figure}

Nous analysons la Théorie de la Grande Jauge basée sur l'action de Wilson (LGT)~\cite{Wilson:1974sk} représentée par $S=-\frac{N_c}{\lambda}\sum_{P} {\mathrm{Re}}\,\mathrm{tr} U_P$, avec $U_P$ désignant le produit de quatre variables de lien unitaires autour de la plaquette $P$. En considérant la limite 't~Hooft ($N_c\to\infty$), l'accent principal est mis sur les Moyennes de Boucle de Wilson (WAs) : $W[C]=\langle\frac{\mathrm{tr}}{N_c}\prod_{l\in C} U_l\rangle$, où le produit matriciel traverse les variables de lien à l'intérieur de la boucle de réseau $C$. Les WAs adhèrent aux Équations de Boucle Makeenko-Migdal (LEs)\cite{Makeenko:1979pb}, ou équations Schwinger-Dyson, qui englobent l'invariance de mesure par rapport aux décalages de groupe, de sorte que $U_l\to U_l(1+i\epsilon)$.\footnote{L'avancement récent implique l'utilisation des équations de boucle pour traiter les problèmes de turbulence.\cite{Migdal:2022bka, Migdal:2023ppb}} Les LEs sont représentées schématiquement :

\begin{equation}\label{MMLE}
\sum_{\nu\perp\mu}\left(W[C_{l_\mu}\!\!\cdot\overrightarrow{\delta C^{\nu}_{l_\mu}}]-W[C_{l_\mu}\!\!\cdot\overleftarrow{\delta C^{\nu}_{l_\mu}}]\right)=\lambda\sum_{\underset{l^\prime\sim l}{l^\prime\in C}}\,\epsilon _{ll^\prime}W[C_{ll^\prime}]\,\,W[C_{{l^\prime l}}]
\end{equation}
avec le LHS indiquant l'action de l'opérateur de boucle sur le lien $l_\mu$, et le RHS signifiant la division du contour $C\to C_{ll^\prime}\cdot C_{l^\prime l}$, comme le montre la Figure~\ref{fig:LE}.

\begin{figure}
\centering
\includegraphics[width=.75\textwidth]{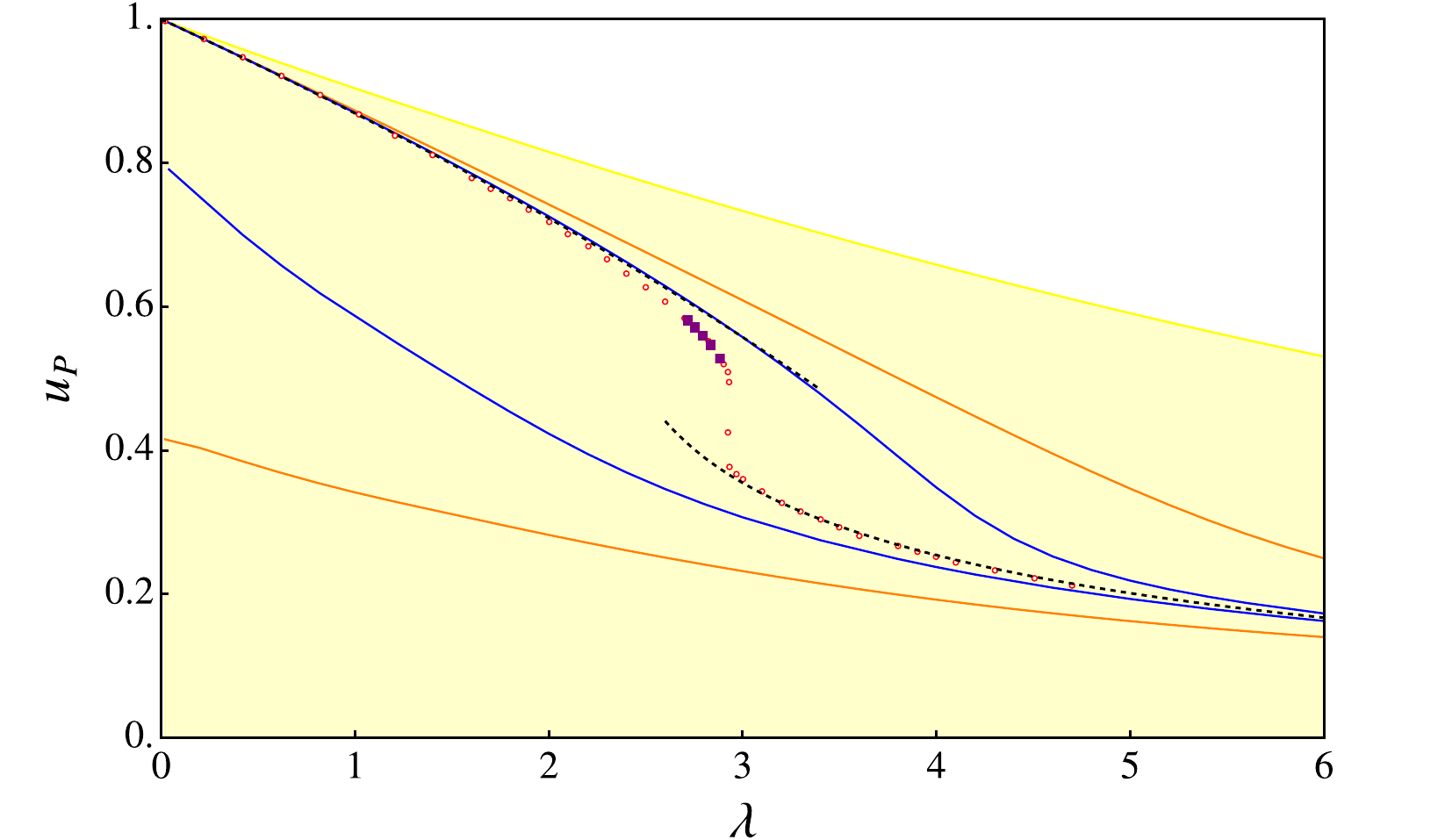}
\caption{La figure présente nos résultats de bootstrap pour les limites supérieures et inférieures de la moyenne de plaquette en \(4D\) LGT. Les domaines pour \(L_{\mathrm{max}}=8, 12, 16\) sont respectivement représentés en jaune, orange et bleu. Les cercles rouges représentent les données de Monte Carlo (MC) pour \(SU(10)\) LGT, avec 5 carrés violets indiquant les résultats pour \(SU(12)\). Les lignes en pointillés supérieure et inférieure signifient respectivement la théorie de perturbation à 3 boucles~\cite{Alles:1998is} et l'expansion de couplage fort~\cite{Drouffe:1983fv}.}
\label{fig:plaquette4D}
\end{figure}

Dans la plupart des cas, le système d'équations de boucle \eqref{MMLE} donne plus de variables de boucle que d'équations de boucle indépendantes. Comme le suggèrent Anderson et al. \cite{Anderson:2016rcw}, la positivité de $\langle  \mathcal{O}^\dagger \mathcal{O}\rangle$ peut être utilisée pour contraindre les quantités dynamiques pertinentes. Nous avons considérablement avancé cette méthode en intégrant les améliorations suivantes :
\begin{enumerate}
    \item Nous prenons en compte les équations de boucle Back-track, qui seront discutées plus en détail ultérieurement.
    \item La positivité de la réflexion du système de réseau est considérée.
    \item La symétrie de réseau est utilisée pour réduire les conditions de positivité.
    \item Nous employons la méthode de relaxation de grande $N$ proposée par les résultats précédents \cite{Kazakov:2021lel}.
\end{enumerate}

Ces améliorations nous ont permis de dériver une limite numérique impressionnante sur la moyenne d'une plaquette $u_P=\frac{1}{N_c}\langle\mathrm{tr} U_P\rangle$. Comme le montre la Fig.~\ref{fig:plaquette4D}, les limites de bootstrap pour \(u_P\) pour \(L_{\mathrm{max}}=8,12,16\) montrent un raffinement rapide à mesure que les coupures augmentent. La limite supérieure encapsule efficacement la phase de boucle de Wilson physiquement significative et reproduit de manière fiable la Théorie de Perturbation à 3 boucles sur une vaste plage de couplage, dépassant même le point de transition de phase. Une comparaison avec les données de Monte Carlo, cependant, indique une marge d'amélioration, notamment dans l'intervalle \((2.4, \,2.8)\) où les données divergent de la Théorie de Perturbation. Cet intervalle, provenant de \cite{Athenodorou:2021qvs}, a été utilisé pour calculer les masses et la tension de la corde. Une amélioration notable est attendue en atteignant \(L_{\mathrm{max}}=20 \text{ ou } 24\), bien que cela nécessitera d'importantes ressources informatiques.
\subsection{Bootstrap conforme}

Les recherches académiques de l'auteur pendant sa thèse de doctorat se sont également étendues au bootstrap conforme.

Des études récentes ont souligné les contraintes strictes que les données CFT, définissant les corr\'elateurs d'opérateurs locaux, doivent respecter \cite{Poland:2018epd}. Ces restrictions ne se limitent pas à encadrer l'espace théorique CFT, mais positionnent également de manière intrigante des théories saillantes à la lisière du domaine autorisé \cite{El-Showk:2012cjh}. La communauté du bootstrap conforme a jusqu'à présent enquêté sur cet espace, explorant principalement deux directions de contraintes sur un ensemble spécifique de corr\'elateurs à quatre points. Premièrement, nous maximisons l'écart dans les dimensions d'échelle du premier opérateur dans l'Expansion du Produit d'Opérateur (OPE) \cite{Rattazzi:2008pe}. Deuxièmement, nous limitons le coefficient OPE d'opérateurs spécifiques au sein de ces corr\'elateurs \cite{Rattazzi:2010gj}. Ces limites sont établies à l'aide de fonctionnels numériques ou analytiques, offrant des perspectives doubles sur le paysage CFT. Ce travail a proposé de nouvelles directions en limitant les valeurs des corr\'elateurs CFT, un problème naturel mais inexploré. Cette approche, similaire à la méthode de minimisation ou de maximisation des valeurs de la matrice $S$, devrait révéler des territoires inconnus dans l'espace CFT. Le problème de la matrice $S$, connu pour produire des limites saturables par des théories passionnantes, est étroitement lié aux corrélateurs CFT (ou à leurs limites) \cite{Paulos:2016fap}, suggérant un potentiel équivalent pour le problème CFT correspondant.

\begin{figure}[ht]
  \centering
	\includegraphics[width=15cm]{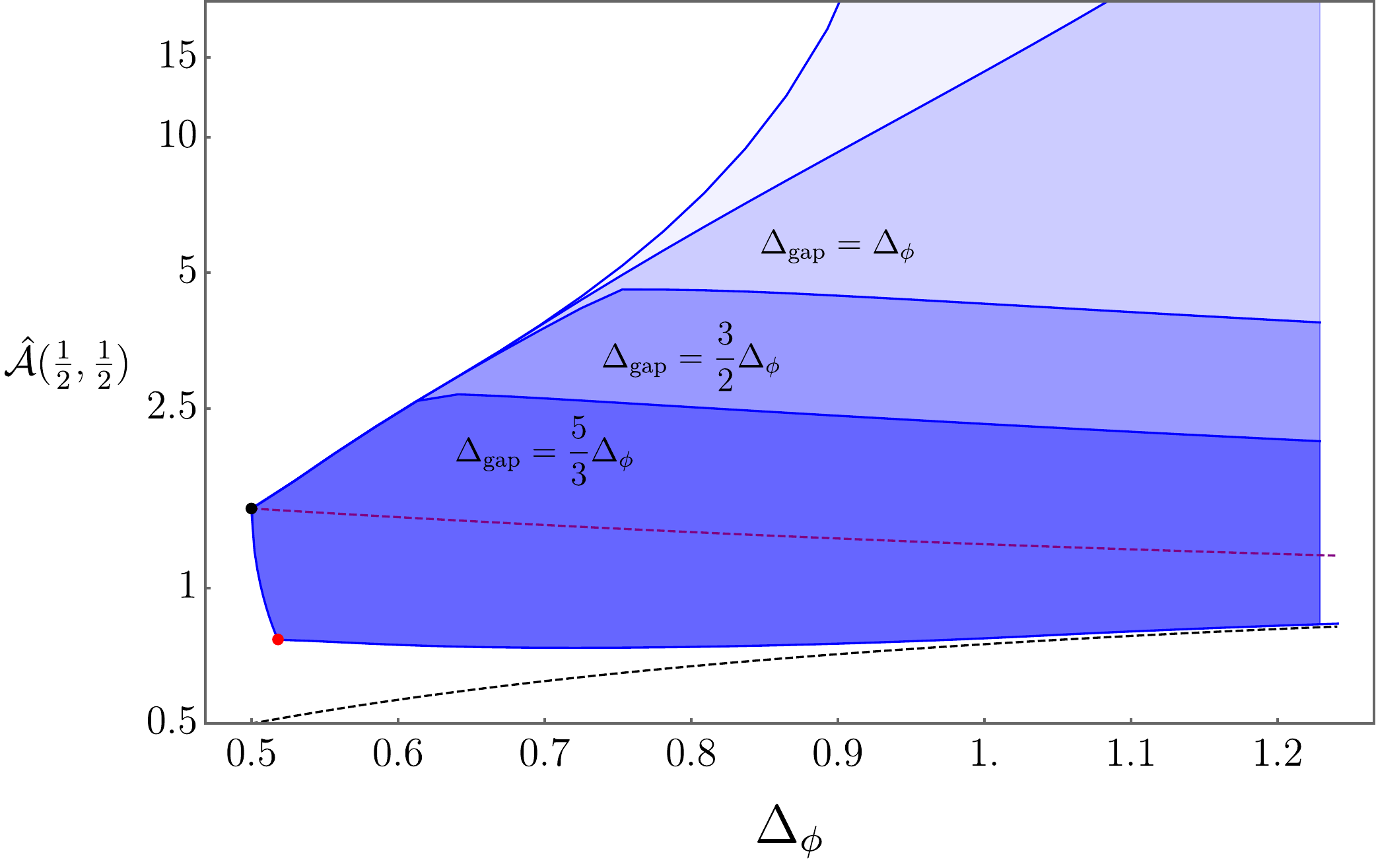}
\caption{\label{allowed}Limites sur les valeurs des corr\'elateurs CFT 3d, avec $\hat{\mathcal A}(z,\bar z)=(z\bar z)^{\Delta_\phi}\mathcal G(z,\bar z)-1$. La région ombragée représente les valeurs qu'un tel corr\'elateur peut prendre au point symétrique de croisement $z=\bar{z}=\frac 12$. Les lignes en pointillés à l'intérieur de la région autorisée correspondent aux valeurs maximales autorisées en supposant un écart. La limite supérieure tend vers $\infty$ près de $\Delta_\phi=1$. La ligne en pointillés à l'extérieur de la région autorisée est le corr\'elateur de fermions libres généralisés 1d, qui fournit une limite inférieure (non optimale). La ligne en pointillés à l'intérieur de la région autorisée est le corr\'elateur de bosons libres généralisés, qui fournit une limite supérieure pour $\Delta_{\mathrm{gap}}=2\Delta_\phi$. Les points noirs et rouges sont respectivement les valeurs de la théorie libre et du modèle d'Ising 3d.}
\end{figure}

Nos résultats numériques, illustrés dans la Figure \ref{allowed}, montrent la plage autorisée du corr\'elateur à quatre points au point de symétrie de croisement $z=\bar z=\frac 12$ en fonction de $\Df$, et les limites le long de $z=\bar z$ pour différentes valeurs de $\Df$.

La Figure \ref{allowed} révèle une limite supérieure du corr\'elateur pour certaines gammes de $\Df$, indépendamment des hypothèses de spectre. Imposer des écarts dans le secteur scalaire affine cette limite, le corr\'elateur de bosons libres généralisés se rapprochant étroitement de la limite supérieure pour un écart de $2\Df$, en accord avec la limite exacte dans~\cite{Paulos:2020zxx}\footnote{L'argument s'appuie sur le fonctionnel analytique, une construction théorique spécifique en théorie des champs conformes avec une base historique étendue~\cite{Afkhami-Jeddi:2020hde, Carmi:2019cub, Caron-Huot:2020adz, Caron-Huot:2022sdy, Dey:2016mcs, Dey:2017fab, El-Showk:2012vjm, El-Showk:2016mxr, Ferrero:2019luz, Ghosh:2021ruh, Ghosh:2023lwe, Ghosh:2023onl, Giombi:2020xah, Gopakumar:2016cpb, Gopakumar:2016wkt, Gopakumar:2021dvg, Hartman:2019pcd, Hartman:2022zik, Kaviraj:2018tfd, Kaviraj:2021cvq, Li:2023whn, Mazac:2016qev, Mazac:2018biw, Mazac:2018mdx, Mazac:2018ycv, Mazac:2019shk, Paulos:2019fkw, Paulos:2019gtx, Paulos:2020zxx, Qiao:2017lkv, Trinh:2021mll}.}. Pour $\Df=1$, la limite diverge en raison de la solution de l'équation de croisement unitaire dépourvue d'identité pour $\Df\geq d-2$. Cela conduit à l'absence de limite sans une hypothèse d'écart de spectre, nécessitant l'écart $\Delta_g > \Df/2$.

À $\Df=1/2$, le point de théorie libre, les limites supérieure et inférieure convergent avec la valeur de la théorie libre, comme on pouvait s'y attendre pour le seul corr\'elateur CFT avec cette dimension. La limite inférieure est également congruente avec la limite exacte déterminée par la solution de fermion libre généralisé 1d, bien que plus forte en raison des contraintes de l'équation de croisement 3d.

La limite inférieure affiche une cassure prononcée à $\Df\sim 0.518149$, la dimension du champ de spin dans le CFT d'Ising 3d. Cela renforce notre argument reliant la minimisation du corr\'elateur à la maximisation de l'écart, car ce dernier conduit au modèle d'Ising 3d à la valeur précise de $\Df$ :
\begin{equation}
    \Delta_g^{\mbox{\tiny min}}\sim \Delta_{\mbox{\tiny gapmax}}
\end{equation}

\begin{singlespace}
\setlength\labelalphawidth{0em}
\small\printbibliography[heading=bibintoc,title=Reference]
\end{singlespace}

\end{document}